\documentclass[11pt]{article}
\pdfoutput=1
\usepackage{amsmath,amssymb}
\usepackage{hyperref}
\usepackage{graphicx}
\newcommand{\ft}[2]{{\textstyle\frac{#1}{#2}}}

\renewcommand{\Im}{\operatorname{Im}}

\newsavebox{\uuunit}
\sbox{\uuunit}
    {\setlength{\unitlength}{0.825em}
     \begin{picture}(0.6,0.7)
        \thinlines
        \put(0,0){\line(1,0){0.5}}
        \put(0.15,0){\line(0,1){0.7}}
        \put(0.35,0){\line(0,1){0.8}}
       \multiput(0.3,0.8)(-0.04,-0.02){10}{\rule{0.5pt}{0.5pt}}
     \end {picture}}

\def\beq{\begin{equation}}
\def\bee{\begin{equation}}
\def\eeq{\end{equation}}
\def\bea{\begin{eqnarray}}
\def\eea{\end{eqnarray}}
\def\bd{\begin{displaymath}}
\def\ed{\end{displaymath}}

\numberwithin{equation}{section}

\begin{document}

\thispagestyle{empty}
{}

\hfill LMU-ASC 76/13\\[-3ex]

\hfill MPP-2013-284 \\
\vskip 6mm

\vskip -3mm
\begin{center}
{\bf\LARGE
\vskip - 1cm
Quantum corrections to extremal \\
black brane solutions \\ [2mm]
}

\vspace{10mm}

{\large
{\bf Susanne Barisch-Dick$^{*\times}$,}
{\bf Gabriel Lopes Cardoso$^\dagger$,} \\ \vskip 3mm
{\bf Michael Haack$^{*}$,}
{\bf \'Alvaro V\'eliz-Osorio$^{^+}$}

\vspace{1cm}

{\it $^*$
 Arnold Sommerfeld Center for Theoretical Physics \\ [1mm]
Ludwig-Maximilians-Universit\"at M\"unchen \\ [1mm] 
Theresienstrasse 37, 80333 M\"unchen, Germany \\ [5mm] 
\it $^\times$
Max-Planck-Institut f\"ur Physik \\ [1mm]
F\"ohringer Ring 6, 80805 M\"unchen, Germany \\ [5mm] 
$^\dagger$
Center for Mathematical Analysis, Geometry, and Dynamical Systems\\ [1mm] 
Departamento de Matem\'atica and LARSyS, 
Instituto Superior T\'ecnico
\\ [1mm]
1049-001 Lisboa, Portugal \\ [2mm] 
$^+$
Departamento de F\'isica,
Instituto Superior T\'ecnico
\\ [1mm]
1049-001 Lisboa, Portugal \\ [2mm] 
}}

\vspace{5mm}

\end{center}
\vspace{5mm}

\begin{center}
{\bf ABSTRACT}\\
\end{center}

We discuss quantum corrections to extremal black brane solutions in $N=2$ $U(1)$ gauged supergravity
in four dimensions. We consider modifications due to a certain class of higher-derivative
terms as well as perturbative corrections to the prepotential.
We use the entropy function formalism to assess the impact of these corrections on singular brane solutions
and we give a few examples. We then use first-order flow equations to construct solutions
that interpolate between quantum corrected fixed points of the associated potentials.

\clearpage
\setcounter{page}{1}

\tableofcontents

\section{Introduction}

Gauged supergravity in four dimensions allows for extremal solutions that have Killing horizons with vanishing entropy density \cite{Goldstein:2009cv}.
These solutions do, however, generically suffer from singularities due to the presence
of tidal forces in the near-horizon region \cite{Kachru:2008yh, Hartnoll:2009sz, Charmousis:2010zz, Copsey:2010ya, Horowitz:2011gh}.\footnote{There are exceptions to this, though, cf.\ \cite{Blau:2009gd, Shaghoulian:2011aa, Lei:2013apa}.}  The associated near-horizon geometry is not 
described by
an $AdS_2 \times \mathbb{R}^2$ line element, 
but instead takes a different form.  It may, for instance, be
of Lifshitz type \cite{Kachru:2008yh}, of hyperscaling violating type \cite{Huijse:2011ef} or 
it may describe an $\eta$-geometry 
\cite{Hartnoll:2012wm}.  There turn out to be various ways to regularize these geometries, for instance by adding electric/magnetic charges
\cite{Kundu:2012jn,Donos:2012yi} or by taking quantum corrections into account 
\cite{Harrison:2012vy,Bhattacharya:2012zu,Knodel:2013fua}.
In both cases the near-horizon geometry of the solution gets modified to
an $AdS_2 \times \mathbb{R}^2$ geometry. This in turn implies that the regularized solution will
have non-vanishing entropy density.\footnote{It is well known that the resulting infrared $AdS_2 \times \mathbb{R}^2$ geometries are often unstable as well, suffering from spatially modulated instabilities, cf.\ \cite{Donos:2011qt, Donos:2011pn, Cremonini:2012ir, Iizuka:2013ag, Donos:2013gda, Cremonini:2013epa}. We will not analyze this kind of instability in the following.}

In this paper we consider quantum corrections to extremal solutions
in $N=2$ gauged
supergravity with $U(1)$ Fayet-Iliopoulos gauging in four dimensions ($N=2 \; U(1)$ gauged supergravity for short). 
This is a step towards a string theory embedding of the proposals to regularize the singular brane solutions by quantum corrections.  
The corrections we consider are of two different types.  They either represent
quantum corrections 
to the prepotential of $N=2$ supergravity, or
they represent higher-derivative corrections proportional to the square of the Weyl tensor.\footnote{We sometimes refer to both of these kinds of corrections as {\it quantum corrections}, even though the higher derivative corrections can arise at tree level in the genus expansion of string theory. However, they do correspond to quantum corrections in the world-sheet theory and it is in this sense that we also refer to them as quantum corrections.}
Analogous effects have been explored in \cite{Harrison:2012vy,Bhattacharya:2012zu,Knodel:2013fua}
in the context of Einstein-Maxwell-dilaton systems.

One way to study the impact of quantum corrections on extremal brane solutions is to study the 
interpolating solution that is obtained by solving the associated first-order flow equations.
First-order flow equations for extremal solutions to $N=2$ $U(1)$ gauged
supergravity in four dimensions were first studied in \cite{Cacciatori:2009iz,Dall'Agata:2010gj,Hristov:2010ri} and reformulated in
terms of homogeneous coordinates in \cite{Barisch:2011ui}.  We will use the latter approach to study the effect of
quantum corrections that are encoded in the prepotential.  On the other hand, if we choose to focus on the near-horizon geometry
of the regularized solutions, the impact of the quantum corrections may also be assessed by using Sen's
entropy function formalism \cite{Sen:2005wa}.  This formalism is amenable to the inclusion of corrections due
to higher-derivative terms, and was explored in the context of extremal black holes in $N=2$ ungauged
supergravity in \cite{Sahoo:2006rp,Cardoso:2006xz}.  Here, we will apply it to $N=2$ $U(1)$ gauged
supergravity in the presence of higher-derivative corrections proportional to the square of the Weyl tensor.

We will begin by deriving the entropy function for extremal black branes in $N=2$ $U(1)$ gauged
supergravity in the presence of the aforementioned higher-derivative terms.  To this end, we adapt
the results of \cite{Sahoo:2006rp,Cardoso:2006xz} to the case at hand. Extremizing the entropy function with respect to the
various fields yields a set of attractor equations whose solution describes the near-horizon solution of
an extremal, not necessarily supersymmetric black brane.  They take a complicated form that simplifies
substantially when restricting to supersymmetric black branes.  We give the form of these attractor equations
with and without higher-derivative terms, and we discuss a few examples, which includes a non-supersymmetric one.
The examples we give describe $AdS_2$ solutions that only exist because of the presence of quantum corrections.

Having constructed $AdS_2$ solutions, we turn to interpolating solutions that interpolate between $AdS_2$ and
$AdS_4$ solutions.  We switch off higher-derivative terms and 
use the formalism of first-order flow equations to construct these interpolating solutions.
We discuss examples where both end points of the flow only exist due to quantum corrections to the prepotential.

\section{The entropy function for extremal black branes}

Extremal black brane solutions with non-vanishing entropy density are solutions which are
supported by scalar fields that are subjected to the attractor mechanism.  
When focussing on the near-horizon
region, the associated attractor equations can be efficiently derived by extremizing Sen's entropy
function \cite{Sen:2005wa}.  
The entropy function framework offers the additional advantage that higher-derivative corrections to the entropy density can be dealt with 
in an efficient manner.

The entropy function formalism relies on the existence of an $AdS_2$ factor in the near-horizon
geometry, but not on supersymmetry.  Thus, the attractor equations derived from the entropy
function formalism will be more general than those derived in the supersymmetric context.  In the following, we derive the attractor equations
for extremal black branes in 
$N=2 \; U(1)$ gauged supergravity 
in the presence of a certain 
class of higher-derivative interactions.  In the absence of the latter, we obtain attractor equations
that encompass those derived in the supersymmetric context 
\cite{Cacciatori:2009iz,Dall'Agata:2010gj,Hristov:2010ri,Barisch:2011ui}.  We give an example of 
a solution that is not supersymmetric.  In the presence of higher-curvature interactions, the
resulting attractor equations are more complicated than their counterparts of the ungauged case.
In order to display the differences between the attractor equations in the gauged and in the ungauged case, we introduce a parameter $k$, related to the curvature of the spatial cross section
of the Killing horizon, that takes the value $k=0$ in the black brane case, and the value $k=1$
in the black hole case.

\subsection{Entropy function}


In the following, we compute the entropy function for extremal black brane solutions
in $N=2$ gauged supergravity with $U(1)$ Fayet-Iliopoulos gauging.
The associated supergravity Lagrangian contains complex scalar fields $X^I$ (with $I = 0, \dots, n$)
that reside in $N=2$ vector multiplets.
We allow for the presence of a class of higher-derivative
terms, namely terms that are proportional to the square of the Weyl tensor.  These so-called $F$-terms
play an important role in $N=2$ string compactifications, 
and they can be dealt with in a systematic fashion by using 
the superconformal approach to supergravity \cite{deWit:1979ug,deWit:1980tn,deWit:1984pk,deWit:1984px}.
In ungauged supergravity, the coupling of the vector multiplets to the Weyl multiplet is 
encoded in a holomorphic function  $F(X, \hat A)$ that is homogeneous of degree two, i.e. $ F(\lambda X, \lambda^2 \, \hat{A}) = \lambda^2 \, F(X, \hat{A})$. Here
$\hat{A} $ denotes the lowest component of the square of the Weyl superfield.
We will assume that in $N=2 \; U(1)$
gauged supergravity these higher-derivative terms are encoded  in the Lagrangian 
through the function $F(X, \hat A)$,  as in the ungauged case. 
Thus, the Lagrangian we will consider is 
\begin{equation}
L= L_u - g^2 \, e^{- 2 \cal K} \, V\;,
\label{mod-lag-flux}
\end{equation}
where $L_u$ denotes the bosonic part of the Lagrangian of $N=2$ ungauged supergravity
with higher-derivative terms \cite{LopesCardoso:2000qm}, and $V $ denotes the flux potential 
\begin{equation}
V= N^{IJ} \hat{h}_I \bar{\hat{h}}_J  - 2 e^{\cal K} \,|W|^2 
\;,
\end{equation}
where
\begin{eqnarray}
N_{IJ} &=& - i \left(F_{IJ} - {\bar F}_{I J} \right) \;, \nonumber\\
e^{- \cal K} &=& i \left( {\bar X}^I F_I - X^I {\bar F}_I  \right) \;, \nonumber\\
\hat{h}_I &=& h_I - F_{IK} h^K \;, \nonumber\\
W &=& h^I F_I - h_I X^I \;.
\label{def-vq}
\end{eqnarray}
Here, $F_I = \partial F(X, \hat{A})/ \partial X^I$ and $F_{IJ} = \partial^2 F(X, \hat{A})/ \partial X^I
\partial X^J$. The $(h_I, h^I)$ denote electric/magnetic fluxes. 
Observe that $V$ is defined in terms of  $F(X, \hat A)$, and that it constitutes 
a symplectic function. The presence of the factor $e^{\cal K} $ ensures that $V$
is invariant under scalings $(X^I, \hat A) \rightarrow (\lambda X^I, \lambda^2 \hat A)$.
In the absence of higher-derivative terms (in which case $F = F(X)$) and in the Poincar\'e
frame (where $e^{-\cal K} =1$),
$V$ reduces to the standard form of the flux potential in  $N=2 \; U(1)$ gauged
supergravity.

The Lagrangian $L_u$ consists of various parts.  One part describes the couplings 
of $N=2$
vector multiplets to supergravity and to the square of the Weyl multiplet, as mentioned above.
Another part describes a hyper multiplet that acts as a 
compensating supermultiplet.  Additional hyper multiplets may be coupled as well, but they will only play a 
passive role in
the following.
The hyper multiplets give rise to the hyper-K\"ahler potential $\chi$.  This field couples to
a real scalar field $D$ that belongs to the Weyl multiplet.

Let us evaluate the Lagrangian \eqref{mod-lag-flux} in an 
$AdS_2$ background,
\begin{eqnarray}
ds^2 = v_1 \left( - r^2 dt^2 + \frac{dr^2}{r^2} \right) + v_2 \, d \Omega^2_k \;,
\label{line-k}
\end{eqnarray}
where $d \Omega^2_k$ denotes the line element of a two-dimensional space of
constant curvature, either flat space ($k = 0$) or
a unit two-sphere $S^2$ ($k = 1$).  Even though we will be interested in extremal black branes,
and hence in the case $k=0$, we will carry $k$ along so as to be able
to compare with the attractor equations for extremal black holes in ungauged supergravity,
which necessarily have $k=1$.  
The background \eqref{line-k} will be supported by electric fields $F_{rt}{}^I = e^I$, magnetic
charges $p^I$ as well as electric and magnetic fluxes $(h_I, h^I)$.
We will consider solutions that have the symmetries of the
line element \eqref{line-k}.  
We follow the exposition of \cite{Cardoso:2006xz}
and adapt the steps given there to the background  \eqref{line-k}.  

In this background, the fields $e^I, X^I, \hat{A}, D, \chi$ take constant values, and the 
Lagrangian $L_u$ will depend on constant parameters $v_1, v_2, e^I, p^I, X^I, 
w, D, \chi$,
where $\hat{A}  = - 4 w^2$ \cite{Sahoo:2006rp,Cardoso:2006xz}.
Since $ \sqrt{-g} \, L_u$ is derived in the superconformal framework, 
it is invariant under rescalings with a complex parameter $\Lambda$, namely \cite{Cardoso:2006xz},
\begin{eqnarray}
  \label{eq:scale-invariance}
  v_{1,2} \to \vert\Lambda\vert^{-2} v_{1,2} \,,\quad
  w \to \bar \Lambda  w \,,\quad
  D \to \vert\Lambda\vert^{2} D \,,\quad
  X^I \to \bar \Lambda X^I \,,\quad 
  \chi\to \vert \Lambda\vert^2 \chi \,,
\end{eqnarray}
while $e^I$ and $p^I$ are invariant under this
scale transformation (and so are the fluxes).  The presence of the factor $e^{-2 \cal{K}}$ in \eqref{mod-lag-flux}
ensures that the 
reduced  Lagrangian $\sqrt{-g} \, L$
will be invariant under this transformation,
and therefore it is natural to express it in terms of scale invariant variables, which
may be chosen as follows \cite{Cardoso:2006xz},
\begin{eqnarray}
  \label{eq:rescaledX}
  && 
  Y^I = \ft14 v_2 \, {\bar w} \, X^I \,,\quad
  \Upsilon = \ft{1}{16} v_2^2 \, {\bar w}^2 \, {\hat A} = - \ft14
  v_2^2 \, \vert w \vert^4 \,, \quad 
  \Xi = \frac{v_1}{v_2} \,,\nonumber\\ 
  &&
  {\tilde D} = v_2 \, \left( D + \tfrac13 R \right)
   \,,\quad 
  \tilde\chi=  v_2 \, \chi \;.  
\end{eqnarray}
Here $R$ denotes the curvature scalar computed in the background \eqref{line-k}
(see appendix \ref{superconf-free}).

Observe that $\Upsilon$ is real and negative, and that
$\sqrt{-\Upsilon}$ and $\Xi$ are real and positive. The potential $V$, when expressed in terms of the rescaled
variables \eqref{eq:rescaledX}, reads
\begin{equation}
V(Y, \bar Y) = N^{IJ} \hat{h}_I \bar{\hat{h}}_J  - 2 \, \frac{|W(Y)|^2}{K(Y, \bar Y)}
\;,
\label{flux-pot}
\end{equation}
where
\begin{eqnarray}
\label{K-W-full}
K(Y, \bar Y) &=& i \left( {\bar Y}^I \, F_I(Y, \Upsilon) - Y^I \, {\bar F}_{I} (\bar Y, \bar \Upsilon) \right) \;,
\nonumber\\
W(Y) &=& h^I F_I(Y, \Upsilon) - h_I Y^I \;.
\end{eqnarray}
Here $F(Y, \Upsilon) $ denotes the rescaled function
$F(X, \hat{A})$, and homogeneity of the function $F(Y,\Upsilon)$
implies 
\begin{equation}
  \label{eq:homo-F}
   F(Y,\Upsilon) = \ft12 Y^I F_I(Y,\Upsilon) + \Upsilon
  F_\Upsilon(Y,\Upsilon)\,,
\end{equation}
where $F_I(Y, \Upsilon) = \partial F(Y, \Upsilon)/\partial Y^I$ and $F_{\Upsilon} = \partial F(Y, \Upsilon)/\partial \Upsilon$.  Further homogeneity relations are listed in appendix 
\ref{homog-rel}.

When imposing the equations of motion for the redefined fields 
$\tilde{D}$ and $\tilde{\chi}$, one finds $\tilde{D} =0$, while $\tilde{\chi}$
gets expressed in
terms of the remaining fields  (see appendix \ref{superconf-free}). Inserting $\tilde{D} =0$ back into 
$L_u$ removes the dependence on $\tilde{\chi}$, since the latter couples
to $\tilde{D}$.  Then, the reduced Lagrangian $\sqrt{-g} \, L$ is expressed in terms of the rescaled parameters 
$Y^I, \Upsilon, \Xi, e^I, p^I$ and the fluxes $(h_I, h^I)$.

The free energy ${\cal F}$ is defined to equal the integral of $\sqrt{-g} \, L$ over a unit cell of the spatial
cross section of the Killing horizon.
Thus, for black branes ($k=0)$, ${\cal F}$ equals $\sqrt{-g} \, L$,
while for black holes ($k=1$) ${\cal F}$ equals the integral of $\sqrt{-g} \, L$ over a unit two-sphere.
The entropy function ${\cal E}$ is defined by the Legendre transform of the free energy ${\cal F}$ with respect
to the electric fields $e^I$, so that
${\cal E} = - {\cal F} - e^I \, q_I$. 
Adapting the results of \cite{Cardoso:2006xz} to the case at hand (see appendix \ref{superconf-free}), we obtain for ${\cal F}$, 
\begin{eqnarray}
  \label{eq:F}
  \tfrac12 \mathcal{F}&=&{}  \tfrac18 N_{IJ} \Big[  \Xi^{-1}
  e^Ie^J - \Xi {p^I} {p^J} \Big] -
  \tfrac14(F_{IJ}+ \bar F_{IJ}) {e^Ip^J}  \nonumber \\
  &&
  + \ft12 i e^I\Big[ F_I + F_{IJ} \bar Y^J  - \mathrm{h.c.}\Big] 
  - \ft12 \Xi\, p^I\Big[ F_I - F_{IJ} \bar Y^J  +  \mathrm{h.c.}\Big] \nonumber\\
 &&
  + \frac{4}{\sqrt{-\Upsilon}} \, K(Y, \bar Y) \,
  (k \, \Xi -1) \nonumber\\
  &&
  + i \,\Xi \Big[ F-Y^IF_I - 2\Upsilon F_\Upsilon+  \tfrac12 \bar
  F_{IJ}Y^IY^J - \mathrm{h.c.} \Big] \nonumber\\
  &&
  + i(F_\Upsilon-\bar F_\Upsilon) \Big [ 32 (k^2 \, \Xi +\Xi^{-1} -2k) - 8 (1+k \, \Xi) \sqrt{-\Upsilon} \Big] \nonumber\\
  &&
  + 32 g^2 \, \Xi  \, \Upsilon^{-1} \, \left[K(Y, \bar Y)\right]^2 \, V(Y, \bar Y) \;,
\end{eqnarray}
while ${\cal E}$ is given by
\begin{eqnarray}
  \label{eq:entropy-total-2}
  \tfrac12  \mathcal{E}  &=& 
   \ft12 \Xi \,\Sigma +\ft12 \Xi \,N^{IJ} (\mathcal{Q}_I-F_{IK}
   \mathcal{P}^K)\, (\mathcal{Q}_J-\bar F_{JL} \mathcal{P}^L) 
   \nonumber\\
   &&
   -\frac{4}{\sqrt{-\Upsilon}} \, K(Y, \bar Y) \,
   (k \, \Xi -1)   \nonumber\\
  &&
  - i (F_\Upsilon-\bar F_\Upsilon) \Big [- 2 \Xi \, \Upsilon + 32
   (k^2 \, \Xi +\Xi^{-1} -2k ) - 8 (1+k \Xi) \sqrt{-\Upsilon} \Big] 
    \nonumber\\ 
  &&
  - 32 g^2 \, \Xi \, \Upsilon^{-1} \, \left[K(Y, \bar Y)\right]^2 \, V(Y, \bar Y) \;.
\end{eqnarray}
To arrive at \eqref{eq:entropy-total-2}, we 
used the homogeneity \eqref{eq:homo-F}
 of the function $F(Y,\Upsilon)$.
The expressions \eqref{eq:F} and \eqref{eq:entropy-total-2} depend 
on $k$, which denotes the curvature of the two-dimensional space
with line element $d \Omega^2_k$.
The quantities 
$\mathcal{Q}_I$, $\mathcal{P}^I$, and $\Sigma$ are defined
by \cite{Cardoso:2006xz}
\begin{eqnarray}
{\cal Q}_I &=& q_I + i \left( F_I - {\bar F}_I \right) \;, \nonumber\\
{\cal P}^I &=& p^I + i \left(Y^I - {\bar Y}^I \right) \;, \nonumber\\
\Sigma &=& - i\left( {\bar Y}^I
  F_I - Y^I {\bar F}_I
  \right) - 2 i \left( \Upsilon F_\Upsilon - \bar \Upsilon
  \bar F_{\Upsilon}\right)
  - q_I   (Y^I+\bar Y^I) + p^I (F_I+\bar F_I) \;.
  \label{comb-QPSIG}
\end{eqnarray}

The entropy function
\eqref{eq:entropy-total-2} depends on the variables $\Xi$, $\Upsilon$
and $Y^ I$, whose values in the near-horizon geometry \eqref{line-k}
are determined by extremizing $\mathcal{E}$.  The resulting equations
are called attractor equations.
In the following, we will discuss the extremization of $\cal E$ with
respect to these variables, first in the absence of higher-derivative terms, and
then with higher-derivative terms. The extremization equations depend on $k$, and
this implies that the attractor equations for black branes in gauged
supergravity (which corresponds to the case $k=0$ and $V \neq 0$) are 
markedly different from those for black holes in ungauged supergravity
(which corresponds to $k=1$ and $V =0$).  When evaluated at the extremum,
the entropy function yields the value of the entropy of the extremal black hole
when $k=1$, and yields the entropy density of the extremal black brane when
$k=0$.

\subsection{Variational equations without higher-derivative terms}
\label{sec:variation-without-R2}

In this subsection, we derive the attractor equations in the absence of 
Weyl interactions. The attractor equations we obtain apply to extremal,
not necessarily supersymmetric black configurations, and they
simplify considerably when restricting to supersymmetric configurations.

When switching off higher-derivative terms, 
the function $F$ does not any longer depend
on $\Upsilon$, i.e. $F = F(Y)$, and the entropy function \eqref{eq:entropy-total-2}
reduces to
\begin{eqnarray}
  \label{eq:entropy-total-3}
   \tfrac12 \mathcal{E}(Y,\bar Y,\Upsilon,\Xi)  &=& 
   \ft12 \Xi \,\Sigma +\ft12 \Xi \,N^{IJ} (\mathcal{Q}_I-F_{IK}
   \mathcal{P}^K)\, (\mathcal{Q}_J-\bar F_{JL} \mathcal{P}^L) 
   \nonumber\\
   &&
   -\frac{4}{\sqrt{-\Upsilon}} \, K(Y, \bar Y) \, 
   (k \, \Xi -1)  \nonumber\\
&&  -  32 g^2 \, \Xi  \, \Upsilon^{-1} \, \left[K(Y, \bar Y)\right]^2 \, V(Y, \bar Y) \;,
  \end{eqnarray}
where now
\begin{equation}
\Sigma = - i \left( {\bar Y}^I
  F_I - Y^I {\bar F}_I
  \right)   - q_I   (Y^I+\bar Y^I) + p^I (F_I+\bar F_I) \;.
  \label{sigma-ups0}
\end{equation}
Varying $\mathcal{E}$ with respect to $Y^I, \Upsilon, \Xi$ and demanding the vanishing of these variations
results in the following equations.
Varying with respect to $\Upsilon$ gives
\begin{equation}
  \label{eq:eq-Upsilon}
\frac{1 - k \, \Xi}{\Xi}  = - 16 g^2  \, \frac{K(Y, \bar Y)}{
 \sqrt{-\Upsilon}} \,
V(Y, \bar Y)\;,
\end{equation}
where we assumed that $K(Y, \bar Y)$ is non-vanishing.  In the ungauged case ($k=1, V=0$) we obtain $\Xi =1$,
which implies $v_1 = v_2$, whereas in the gauged case ($k=0, V \neq 0$) 
$\Xi$ becomes a non-trivial function of $\Upsilon$ and $Y^I$, namely
\begin{equation}
  \label{eq:U-Y-Ups}
\frac{1}{\Xi}  = - 16 g^2 \, \frac{K(Y, \bar Y)}{
 \sqrt{-\Upsilon}}
V(Y, \bar Y)\;,
\end{equation}
where consistency requires the right hand side of 
 \eqref{eq:U-Y-Ups} to be positive.

  Varying \eqref{eq:entropy-total-3} with respect to $\Xi$ yields,
\begin{eqnarray}
  \label{eq:eq-U}
  &&\Sigma +  \left( {\cal Q}_I - F_{IK} \, {\cal P}^K\right) N^{IJ} 
  \left( {\cal Q}_J - {\bar F}_{JL} \, {\cal P}^L \right)
  - \frac{8 k }{\sqrt{-\Upsilon}} \, K(Y, \bar Y) \nonumber\\ 
&&  -  64 g^2   \, \Upsilon^{-1} \, \left[K(Y, \bar Y)\right]^2 \, V(Y, \bar Y) =0 \;,
  \end{eqnarray}
  which determines the value of $\Upsilon$ in terms of the $Y^I$. This reflects the fact
that in the absence of $R^2$ -interactions, 
the quantity $\Upsilon$ is related to an auxiliary field
in the original Lagrangian whose field equation is algebraic. Again, depending on the case ($k=1, V=0$ or 
$k=0, V \neq 0$) the relation is markedly different.  When $k=0$, we get
\begin{eqnarray}
  && \frac{64 g^2}{\Upsilon} = \frac{
  \Sigma +  \left( {\cal Q}_I - F_{IK} \, {\cal P}^K\right) N^{IJ} 
  \left( {\cal Q}_J - {\bar F}_{JL} \, {\cal P}^L \right) }{\left[ K(Y, \bar Y)\right]^2
     V(Y, \bar Y) } \;.
 \label{value-Ups}
   \end{eqnarray}

Let us explore some of the consequences of \eqref{eq:U-Y-Ups}  and \eqref{value-Ups}
in the Einstein frame, where $e^{\cal K} =1$. 
Using the scaling relations  \eqref{eq:rescaledX}, we infer
\begin{equation}
|w|^2 = - \frac{\Upsilon}{4 K(Y, \bar Y)} \;\;\;,\;\;\; v_2 = \frac{8 K(Y, \bar Y)}{\sqrt{- \Upsilon}} \;,
\label{w,v_2}
\end{equation}
and from \eqref{eq:U-Y-Ups} we obtain
\begin{equation}
v_1 = - \frac{1}{2 g^2 \, V(Y, \bar Y)}\;.
\label{eq:v1}
\end{equation}
This expresses the scale factors $v_1, v_2$ 
in terms of $Y^I$ and $\Upsilon$.  
Inserting \eqref{value-Ups} into \eqref{eq:entropy-total-3} 
yields
\begin{equation}
   \mathcal{E} = \frac{8 K(Y, \bar Y)}{\sqrt{- \Upsilon}} \;,
 \label{entro}
   \end{equation}
which, according to \eqref{w,v_2},
 equals $v_2$, as expected for the black brane entropy density.
Now consider a uniform rescaling of the charges $(q_I, p^I)$, of the fluxes $(h_I, h^I)$ 
and of the variables $Y^I$.  Then, we infer from \eqref{value-Ups} that
$[K(Y, \bar Y)]^2/\Upsilon$ is inert under such a rescaling.  It follows that \eqref{entro} scales with
weight zero, and this implies that when expressing $\mathcal{E}$  in terms of charges and fluxes,
it will be of weight zero in the charges and fluxes. This differs markedly from the case of big black holes
in ungauged supergravity, where  $\mathcal{E}$ scales quadratically in the charges.

Next, consider 
varying 
 the entropy function
\eqref{eq:entropy-total-3} with respect to $Y^I$.  We focus on the black brane case $k=0$, and obtain 
\begin{eqnarray}
{\cal P}^J F_{JI} - {\cal Q}_I 
  + \tfrac1{2} i \left({\cal Q}_K -\bar F_{KM} \,{\cal P}^M
  \right) 
  N^{KP} \, F_{PIJ} \, N^{JL} \left( {\cal Q}_L - {\bar F}_{LN}
  {\cal P}^N \right) + \tfrac12 v_2^2 \,g^2 V_I =0 \;,
\label{atrractor-eqs}
  \end{eqnarray}
  where $V_I = \partial V(Y, \bar Y) / \partial Y^I$, and 
we used \eqref{w,v_2} and \eqref{eq:U-Y-Ups}.
Writing out \eqref{atrractor-eqs} gives
\begin{equation}
q_I - F_{IJ} p^J - N_{IJ} {\bar Y}^J - 
\tfrac12  i \left(q_K -\bar F_{KM} p^M
  \right) 
  N^{KP} \, F_{PIJ} \, N^{JL} \left( q_L - {\bar F}_{LN}
  \,p^N \right) =  \tfrac12 v_2^2 \, g^2 V_I,
  \label{attract-1}
  \end{equation}
where we made use of the special geometry relation
$F_{IJK}Y^K=0$.  Next we 
compute $V_I$,
\begin{equation}
V_I = i N^{KP} F_{PQI} N^{QL} \hat{h}_K \bar{\hat h}_L - \frac{2}{K^2} 
\left({\bar Y}^M N_{MI} \, Y^K \hat{h}_K + K \, \hat{h}_I \right) {\bar Y}^N \bar{\hat{h}}_N 
- N^{KL} F_{IKP} h^P \bar{\hat{h}}_L\; ,
\label{eq:VI}
\end{equation} 
which satisfies $V_I \, Y^I =0$.
Using the expression for $V_I$ as well as the relation $\hat{h}_I = \bar{\hat{h}}_I - i N_{IL} h^L$, we obtain
for \eqref{attract-1},
\begin{eqnarray}
\label{hatQ-att1}
\hat{Q}_I - N_{IJ} \bar{Y}^J - \tfrac12 i F_{IJK} N^{JL} N^{KM} \bar{\hat{Q}}_L  \bar{\hat{Q}}_M
&=& \tfrac12 v_2^2 \, g^2 \left[ i N^{PQ} F_{QMI} N^{MN} \bar{\hat{h}}_P  \bar{\hat{h}}_N \right.
\nonumber\\
&& \left. \qquad
- \frac{2}{K^2} 
\left({\bar Y}^M N_{MI} \, Y^K \hat{h}_K + K \, \hat{h}_I \right) {\bar Y}^N \bar{\hat{h}}_N 
\right] \;, \nonumber\\
\end{eqnarray}
where we introduced the combination
\begin{eqnarray}
\label{hatQhat}
\hat{Q}_I = q_I - F_{IJ} \, p^J \;.
\end{eqnarray}
Finally, we rewrite \eqref{hatQ-att1} as
\begin{eqnarray}
&&\hat{Q}_I + g^2 \, \frac{v_2^2}{K}\,  {\bar Y}^N \bar{\hat{h}}_N \, \hat{h}_I \nonumber\\
&& \qquad 
- \tfrac12 i F_{IJK} N^{JL} N^{KM} \left( \bar{\hat{Q}}_L  \bar{\hat{Q}}_M
+ g^2 \, v_2^2 \, \bar{\hat{h}}_L  \bar{\hat{h}}_M \right) \nonumber\\
&& \qquad \qquad = {\bar Y}^J N_{JI} \left( 1 - g^2 \, \frac{v_2^2}{K^2} \, Y^K \hat{h}_K \, {\bar Y}^N \bar{\hat{h}}_N 
\right) \;.
\label{attrac-2}
\end{eqnarray}
These are the black brane attractor equations for the $Y^I$.  Contracting them with $Y^I$ yields the constraint
\begin{equation}
K(Y, \bar Y) = - \hat{Q}_I Y^I \equiv Z(Y) = \bar{Z}(\bar Y) \;,
\label{K-Z}
\end{equation}
and hence, on the attractor, we obtain from \eqref{sigma-ups0}, 
\begin{equation}
\Sigma = K(Y, \bar Y) \;.
\end{equation}

Next, we relate 
the entropy function \eqref{entro} to the black hole potential which, in the Poincar\'e frame ($e^{\cal K}=1$), takes the form
\begin{equation}
V_{\rm BH} = \left[ N^{IJ} + 2 \, X^I \, {\bar X}^J \right] \hat{Q}_I \bar{\hat{Q}}_J \;,
\end{equation}
as follows.  First we observe that \eqref{entro} can be written as 
\begin{eqnarray}
    \mathcal{E} =
   2 \, \Xi \left[ 2 Z(Y)  + N^{IJ} \hat{Q}_I \bar{\hat{Q}}_J \right]
\end{eqnarray}
by making use of \eqref{eq:U-Y-Ups}, \eqref{value-Ups} and \eqref{K-Z}.
Next, we express $Y^I$ as 
\begin{equation}
Y^I =  {\bar Z}( \bar X) \, X^I \;,
\end{equation}
where 
\begin{equation} \label{Zx}
Z(X) = -  \hat{Q}_I \, X^I \;,
\end{equation}
which is consistent with \eqref{eq:rescaledX} by virtue of \eqref{w,v_2} and \eqref{K-Z}.
This 
yields $Z(Y) = |Z(X)|^2$, which results in
\begin{eqnarray}
    \mathcal{E} =    2 \, \Xi \, V_{\rm BH} \;.
    \label{entro-Vbh}
\end{eqnarray}

Note that the entropy function approach needs to be supplemented by the Hamiltonian constraint,  which imposes the following
restriction on the charges and the fluxes \cite{Dall'Agata:2010gj},
\begin{equation}
q_I \, h^I = p^I \, h_I \;.
\label{hamil}
\end{equation}

Now let us return to the black brane attractor equations \eqref{attrac-2}.
They take a form that is very different from their black hole counterpart in ungauged supergravity 
\cite{Sahoo:2006rp,Cardoso:2006xz}.
The latter simplify substantially when restricting to supersymmetric solutions, in which case they are
given by ${\cal Q}_I = {\cal P}^I =0$.  In the case of gauged supergravity, an analogous simplification
occurs when considering supersymmetric solutions, as follows.  To solve \eqref{attrac-2}, we make the ansatz
\begin{equation}
\hat{Q}_I = g \, e^{i \delta} v_2 \, \hat{h}_I \;,
\label{Y-ansatz}
\end{equation}
which, upon contraction with $Y^I$, results in 
\begin{equation}
Z(Y) = g\, e^{i \delta} v_2 \, W(Y) = g\, e^{- i \delta} v_2 \, \bar{W}(\bar Y) = 
K(Y, \bar Y)\;.
\label{W-bW}
\end{equation}
Inserting these relations into \eqref{attrac-2}  gives
\begin{eqnarray}
\hat{Q}_I -g\,  e^{i \delta} v_2 \, \hat{h}_I 
- \tfrac12 i F_{IJK} N^{JL} N^{KM}  \bar{\hat{h}}_L  \bar{\hat{h}}_M \,g^2 \, 
v_2^2 \left(1 + e^{- 2 i \delta} \right)  = 0 \;.
\label{attrac-3}
\end{eqnarray}
This vanishes provided that
\begin{equation}
e^{- 2 i \delta} = -1 \;.
\label{value-delta}
\end{equation}
Inserting \eqref{value-delta} into \eqref{Y-ansatz} yields the attractor values
derived in \cite{Dall'Agata:2010gj,Barisch:2011ui}.  They apply to supersymmetric solutions \cite{Dall'Agata:2010gj}
as well as to solutions derived from supersymmetric ones by applying a transformation ${\cal S}$ 
\cite{Ceresole:2007wx} to the charges
and to the fluxes. We will refer to \eqref{Y-ansatz} and \eqref{value-delta} as supersymmetric attractor equations, for simplicity.
They 
constitute a simplification compared to the non-supersymmetric
ones based on \eqref{attrac-2}.

Since in \eqref{Y-ansatz} the dependence on $Y^I$ only enters through $F_{IJ}$,
which is homogeneous of degree zero, the $Y^I$ only appear as ratios, i.e.\ as projective
coordinates $z^i = Y^i/Y^0$ with $i = 1, \dots, n$.  The equations \eqref{Y-ansatz}
can thus be viewed as equations that determine the values of the $n$ parameters
$z^i$ and $v_2$ in terms of charges and fluxes.  
Using \eqref{W-bW}, \eqref{value-delta} and \eqref{w,v_2}, we find that the supersymmetric attractor equations 
 can be recast in the form
\begin{eqnarray}
\label{low-ord-attr}
&& \hat{Q}_I  + 64 g^2 \Upsilon^{-1} K {\bar W} \hat{h}_I = 0  \;,\nonumber\\
&& - \Upsilon =  64 g^2   \,|W|^2 \nonumber\\
 &&   \sqrt{- \Upsilon} \, \Xi^{-1} = -  16 g^2  \,
  K  \, V \;, \nonumber\\
  && W = - \bar W \;.
\end{eqnarray}
Note that 
combining \eqref{hamil} with \eqref{W-bW} results in 
$v_2 \, \hat{h}_I N^{IJ} \bar{\hat{h}}_J =0$ \cite{Dall'Agata:2010gj,Barisch:2011ui}.  Taking $v_2 \neq 0$, and inserting 
$\hat{h}_I N^{IJ} \bar{\hat{h}}_J =0$ into \eqref{flux-pot} yields $V = - 2 |W|^2/K$ on a supersymmetric attractor.
Combining this with \eqref{low-ord-attr} gives 
\begin{equation}
\Xi^{-1} = 4 g \, |W(Y)| \;.
\end{equation}

The attractor equations for the scalars, \eqref{Y-ansatz} together with \eqref{value-delta}, can be obtained by
extremizing the effective potential of the associated one-dimensional effective Lagrangian \cite{Dall'Agata:2010gj,Barisch:2011ui}.\footnote{We thank the referee for raising this question.}  The latter is obtained by evaluating the Lagrangian in a static black brane background with line element
\begin{equation}
ds^2 = - e^{2U} dt^2 + e^{-2U} dr^2 + e^{2(\psi - U)} (dx^2 + dy^2)
\end{equation}
which, at the horizon, reduces to the line element  \eqref{line-k} (with $k=0$).  The resulting effective
potential takes the form
\begin{eqnarray}
V_{\rm tot} &=& g^2 \big[ N^{IJ} \partial_I \tilde{W} \partial_{\bar J} \bar{\tilde{W}} - 2 | \tilde{W}|^2 \big]
+ e^{-4(\psi - U)} \big[ N^{IJ} \partial_I \tilde{Z} \partial_{\bar J} \bar{\tilde{Z}} + 2 | \tilde{Z}|^2 \big] \;, 
\end{eqnarray}
where $Z$ and $W$ are given by \eqref{Zx} and \eqref{def-vq}, respectively, with $X^I$ replaced by the rescaled, $U(1)$ invariant
field $\tilde{X}^I = {\bar \varphi} \, X^I$ \cite{Barisch:2011ui}.
Here $\partial_I = \partial/\partial \tilde{X}^I$.
This effective potential can be expressed in terms of a quantity $\Delta$ given by
\begin{equation}
\Delta = e^{2 U} Z(\tilde{X}) - i g \, e^{2 \psi}\, W (\tilde{X}) \;,
\end{equation}
which depends holomorphically on $\tilde{X}^I$.  We then obtain
\begin{eqnarray}
V_{\rm tot} 
= e^{ - 4 \psi} \left[ N^{IJ} \partial_I \Delta \partial_{\bar J} \bar \Delta + \frac12 \partial_U \Delta \partial_U \bar \Delta - \frac12 \partial_\psi \Delta \partial_\psi \bar \Delta \right] + g e^{2(U - \psi)} \,
(q_I h^I - p^I h_I)
\;.
\end{eqnarray}
Now consider varying $V_{\rm tot}$ with respect to the scalar fields $\tilde{X}^I$.
The variation of the first term in the bracket can be set to zero by demanding $\partial_I \Delta =0$,
which in turn implies $\Delta =0$, since $\Delta = \tilde{X}^I \, \partial_I \Delta$ by virtue of special geometry.
The variation of the sum of the second and third terms in the bracket also vanishes when imposing $\partial_I \Delta =0$.
Thus, we obtain an extremum of the potential by demanding $\partial_I \Delta =0$.
If we now take $U$ and $\psi$ to have the form of an $AdS_2$ background \eqref{line-k}, i.e. $e^{2U} = r^2/v_1, e^{2 \psi} = v_2 \,
e^{2U}$ (with a subsequent rescaling $t \rightarrow v_1 t$), we obtain from $\partial_I \Delta =0$, 
\begin{equation}
\hat{Q}_I - i g v_2 \hat{h}_I =0 \;,
\end{equation}
in agreement with \eqref{Y-ansatz} and \eqref{value-delta}. 

In the ungauged case, it is known that extrema of the effective potential correspond to
minima, as long as the metric on the moduli space of physical scalars is positive definite, 
 cf.\ \cite{Ferrara:1997tw}. In the case at hand, it is not obvious that 
 an extremum is a minimum of the effective potential, as we proceed to
 analyze.  To do so,  we have to take into account that the scalar fields $X^I$ are constrained to satisfy $N_{IJ} X^I \bar{{X}}^J = -1$. Expressing the $X^I$ in terms of the physical scalar fields $z^i = X^i/X^0$ $(i=1,\ldots, n)$, we obtain
\begin{eqnarray} 
D_i \left(|Z|^2 + D_k Z g^{k \bar \jmath} {\bar D}_{\bar \jmath} {\bar Z} \right) &=& 2 (D_i Z) \bar{Z} + i C_i{}^{\bar k \bar l} 
{\bar D}_{\bar k} \bar{Z} {\bar D}_{\bar l} \bar{Z}\ , \nonumber\\
D_i \left(- 3 |W|^2 + D_k W g^{k \bar \jmath} {\bar D}_{\bar \jmath} {\bar W} \right) &=& - 2 (D_i W) \bar{W} + i C_i{}^{\bar k \bar l} 
{\bar D}_{\bar k} \bar{W} {\bar D}_{\bar l} \bar{W}\ , \label{DWDZ}
\end{eqnarray}
where $g^{k \bar \jmath}$ is the inverse of the metric on the moduli space of the physical scalars, $C_{ijk}$ is a covariantly holomorphic symmetric tensor and $D_i$ is the covariant derivative with respect to the usual Levi-Civita connection and the K\"ahler connection. In deriving \eqref{DWDZ}, we used the following identities from special geometry (see for instance (3.2) in \cite{Ceresole:1995jg})
\begin{eqnarray}
D_i D_j X^I &=& i C_{ijk} g^{k \bar l} \bar D_{\bar l}  \bar{X}^I\ , \nonumber \\
D_i \bar D_{\bar \jmath} \bar{X}^I &=&  g_{i \bar \jmath}  \bar{X}^I \ , \nonumber \\
D_i \bar{X}^I &=& 0 \ . \label{special_geo_ids}
\end{eqnarray}
Using the identity (see (23) of \cite{deWit:1996ag}) 
\begin{equation}
N^{IJ} = g^{i \bar \jmath} \, D_i X^I \, \bar D_{\bar \jmath} \bar{X}^J  - X^I \, \bar{X}^J \;,
\label{eq:N-id-X}
\end{equation}
one sees that \eqref{DWDZ} and $\Delta =0$ indeed imply $\partial_i V_{\rm tot} =0$, i.e.\ an extremum. 
Using the identities \eqref{special_geo_ids}, one also shows that at the extremum, 
\begin{eqnarray}
D_j D_i V_{\rm tot} &=& D_j \partial_i  V_{\rm tot} = \partial_j \partial_i  V_{\rm tot} = e^{-4(\psi - U)} 4 i C_{ij}{}^{\bar k}  {\bar Z} {\bar D}_{\bar k} {\bar Z}\ , \nonumber \\
{\bar D}_{\bar \jmath} D_i  V_{\rm tot} &=& \partial_{\bar \jmath} \partial_i  V_{\rm tot} = e^{-4(\psi - U)} 4 
 C_i{}^{\bar k \bar l} {\bar C}_{\bar \jmath \bar l}{}^p \, D_p Z {\bar D}_{\bar k} {\bar Z}\ .
\end{eqnarray}
This is markedly different from the ungauged case, where 
$ \partial_j \partial_i  V$ vanishes 
at the extremum and $\partial_{\bar \jmath} \partial_i  V$
is positive definite there \cite{Ferrara:1997tw}. Thus, in the presence of fluxes, a more
detailed analysis is required 
to decide whether an extremum of $V_{\rm tot}$ is actually a minimum. This we leave for future work.

Summarizing, for extremal black brane solutions ($k=0$) the attractor equations for $\Xi, \Upsilon$ and $Y^I$ are
given by \eqref{eq:U-Y-Ups}, \eqref{value-Ups} and \eqref{attrac-2}.  In the supersymmetric case, these become \eqref{low-ord-attr}.
The entropy density is related to the black hole
potential by \eqref{entro-Vbh}.  When expressed in terms of charges and fluxes, it has weight zero under uniform
scalings of the charges and of the fluxes.

Finally, let us consider the free energy \eqref{eq:F}.
Using \eqref{eq:eq-Upsilon} and introducing the combination ${\cal Y}^I = \tfrac12 \left( \Xi^{-1} \, e^I + i p^I \right)$, we obtain 
\begin{eqnarray}
\tfrac12 \,  \mathcal{F}&=&{} \tfrac14  \Xi \Big[  N_{IJ} \left( {\cal Y}^I {\cal Y}^J + \bar{\cal Y}^I \bar{\cal Y}^J \right)
  + i \left(F_{IJ} + {\bar F}_{IJ} \right) \left( {\cal Y}^I {\cal Y}^J - \bar{\cal Y}^I \bar{\cal Y}^J \right)
  \Big] \nonumber \\
  &&
  - \ft12 \Xi \, N_{IJ} \Big[ \left( {\cal Y}^I +  \bar{\cal Y}^I \right) \left( Y^J +  \bar{Y}^J \right) + 
  \left( {\cal Y}^I -  \bar{\cal Y}^I \right) \left( Y^J -  \bar{Y}^J \right) \Big] \nonumber\\
 && + \tfrac12  \Xi \,   N_{IJ} \left( {Y}^I {Y}^J + \bar{Y}^I \bar{Y}^J \right)  \nonumber\\
 &&
   -32 g^2 \, \Xi \,  \Upsilon^{-1} \, [K(Y, \bar Y)]^2 \, 
     V(Y, \bar Y) \;.
\end{eqnarray}
In the absence of fluxes we have $k=1$ and $\Xi=1$, as can be seen from \eqref{eq:eq-Upsilon}.
In this case, BPS solutions satisfy $Y^I = {\cal Y}^I$, and the free energy evaluated on these
solutions equals \cite{Ooguri:2004zv},
\begin{equation}
  \mathcal{F} = - 4 \, {\rm Im} \, F(Y) \;.
\end{equation}
In the presence of fluxes, no analogous simplification occurs.

\subsubsection{Examples}

The attractor equations \eqref{attrac-2} allow for supersymmetric solutions 
as well
as for non-super\-symmetric solutions.  In the following, we give two examples of solutions to the attractor equations.
The first example is non-supersymmetric, while the second example
is supersymmetric, and therefore satisfies \eqref{Y-ansatz}.  The first example is based on the prepotential
\begin{equation}
F(Y) = - \left(Y^1\right)^3/Y^0 + i c \, (Y^0)^2 = i \left(Y^0\right)^2 \left(t^3 + c \right)\;,
\label{t-sugra-c}
\end{equation}
where $t = - i Y^1/Y^0$ and $c< 0$. We take the solution to be supported by a non-vanishing electric charge $q_0$ and
a non-vanishing electric flux $h_1$ satisfying $q_0 \, h_1 < 0$.  Then, we find that the attractor equations
\eqref{eq:U-Y-Ups}, \eqref{value-Ups} and \eqref{attrac-2} can be solved {\it exactly}, with the solution given by
\begin{equation}
t = \beta_1 \, |c|^{1/3} \;,\; \Xi = \beta_2 \, \frac{|c|^{2/3}}{|q_0 h_1|} \;,\; Y^0 = \beta_3 \, \frac{ q_0}{c}
\;,\; \sqrt{- \Upsilon} = \beta_4 \, \frac{|q_0 h_1|}{|c|^{2/3}} \;,
\end{equation}
where the $\beta_i$ denote fixed real constants given by
\begin{eqnarray}
\beta_1= 0.323 \;\;,\;\; \beta_2=1.971 \;\;,\;\; \beta_3=0.234 \;\;,\;\;
\beta_4=2.267 \;.
\end{eqnarray}
Observe that the solution only exists because of the presence of the  $c$-term
in $F(Y)$, and that the modulus $t$ takes a real and positive value that is independent of $(q_0, h_1)$.  
On the solution, $F(Y) \neq 0$ since $\beta_1^3 + 1 \neq 0$.
Using
\eqref{w,v_2} we obtain
\begin{equation}
v_2= - 32 \, \frac{(2\beta_1^3+ 1) \beta_3^2}{\beta_4} \, \frac{q_0}{h_1} \, \frac{1}{|c|^{1/3}}\;,
\end{equation}
which is positive.  In the limit of large $c$, $t$ becomes large, while $v_2$ shrinks to zero.
When embedding a supergravity model of the form \eqref{t-sugra-c} into type II string theory,
requiring a large value of ${\rm Re} \, t$ is necessary in order to neglect worldsheet instanton
contributions to $F(Y)$. However, in type II string theory the term $c$ constitutes an $\alpha'$
correction, and hence a subleading term, while for the above solution both terms in $F(Y)$,
$t^3$ and $c$,
are of similar order (even though there is indeed a small hierarchy as $\beta_1^3 \approx 1/30$).  
Thus, while the above solution constitutes a solution to the supergravity
toy model \eqref{t-sugra-c}, for it to also constitute a solution to a string model would require
taking worldsheet instanton effects into account.

\begin{figure}[t]
   \includegraphics[scale=0.28]{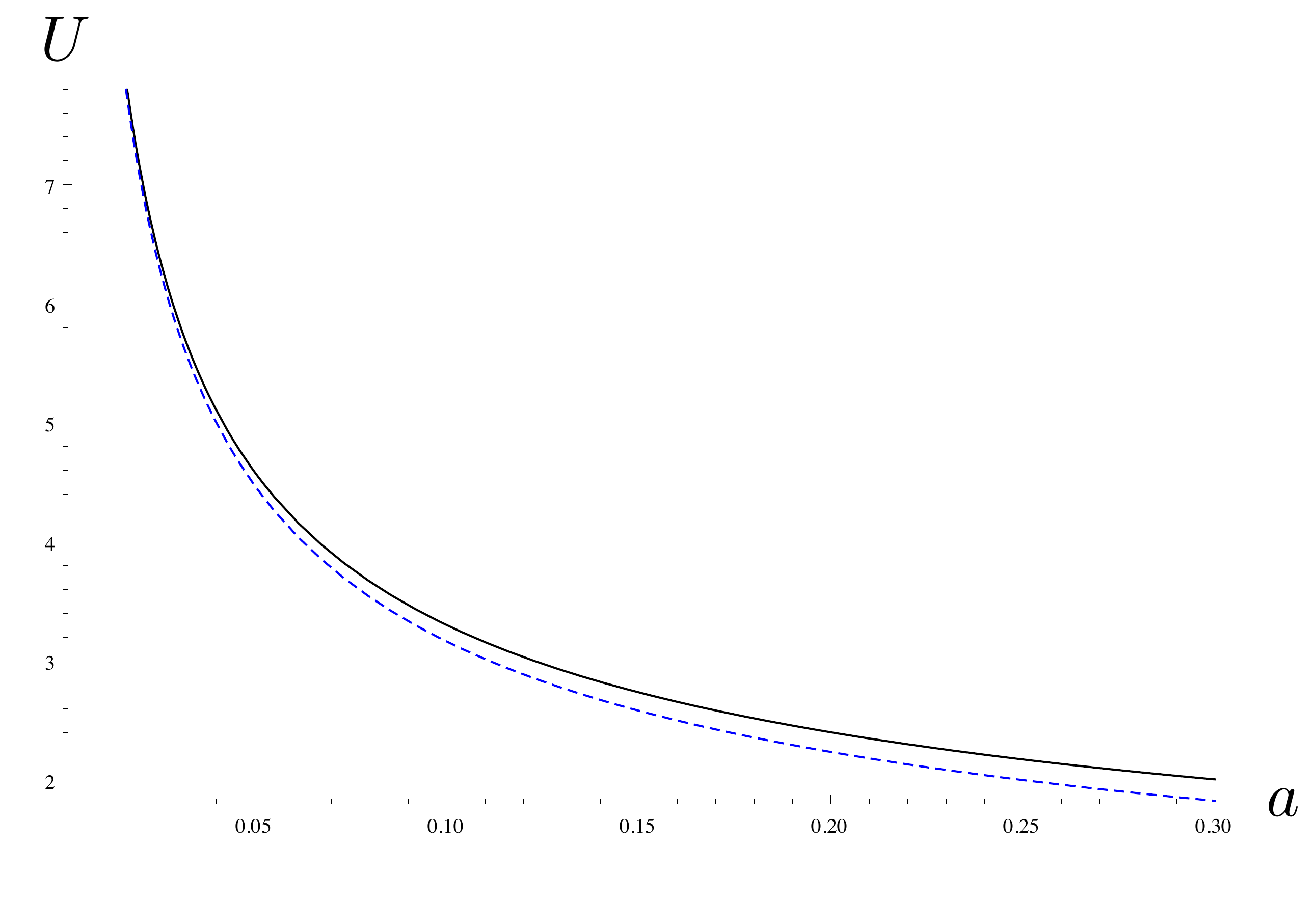} 
   \includegraphics[scale=0.275]{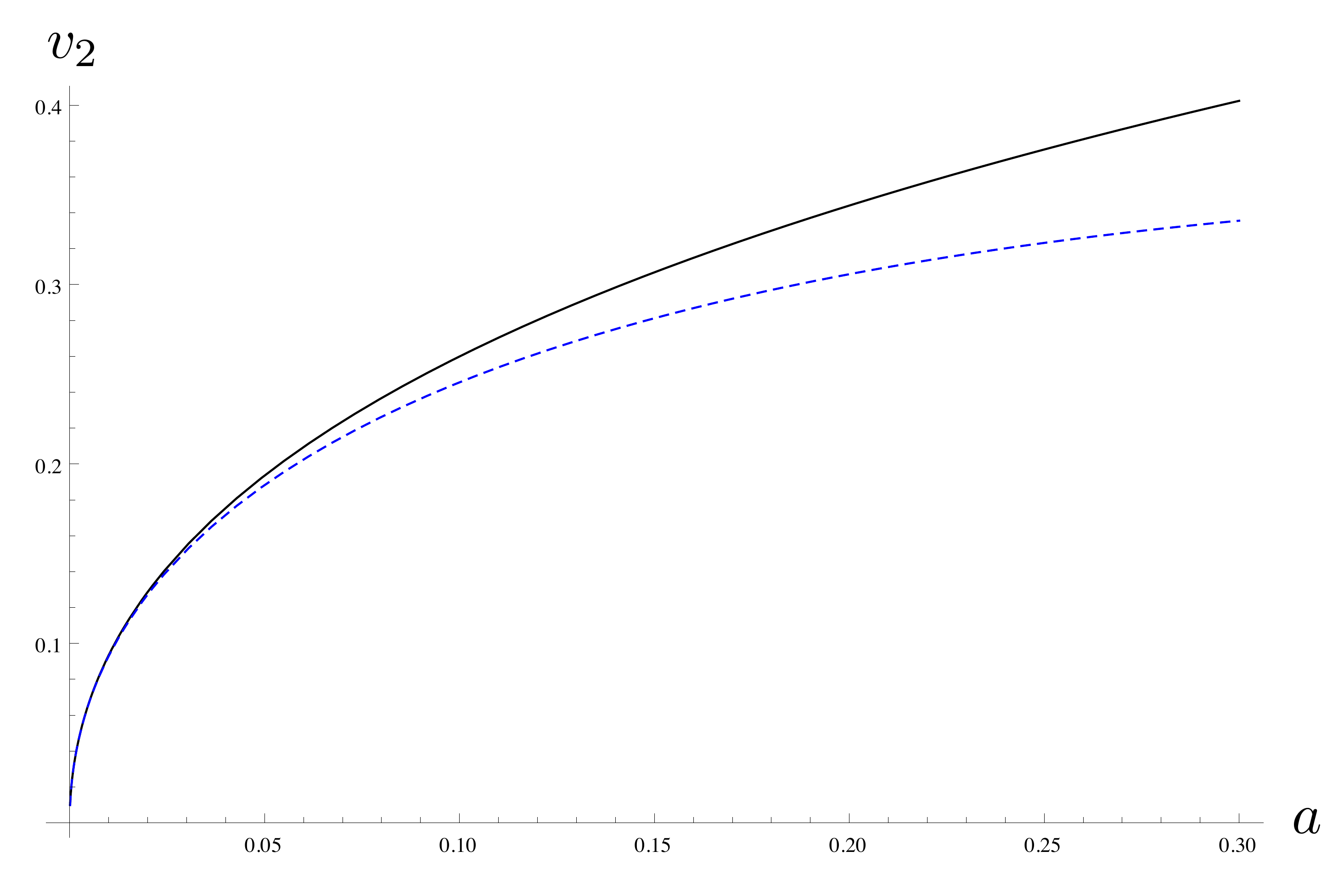}\\
  \caption{Dashed (leading approximation), solid (exact) $(h^0=h_1=h_2=h_3=p^3=1$).}\label{BB}
\end{figure}

\begin{figure}[t]
  \center{\includegraphics[scale=0.4]{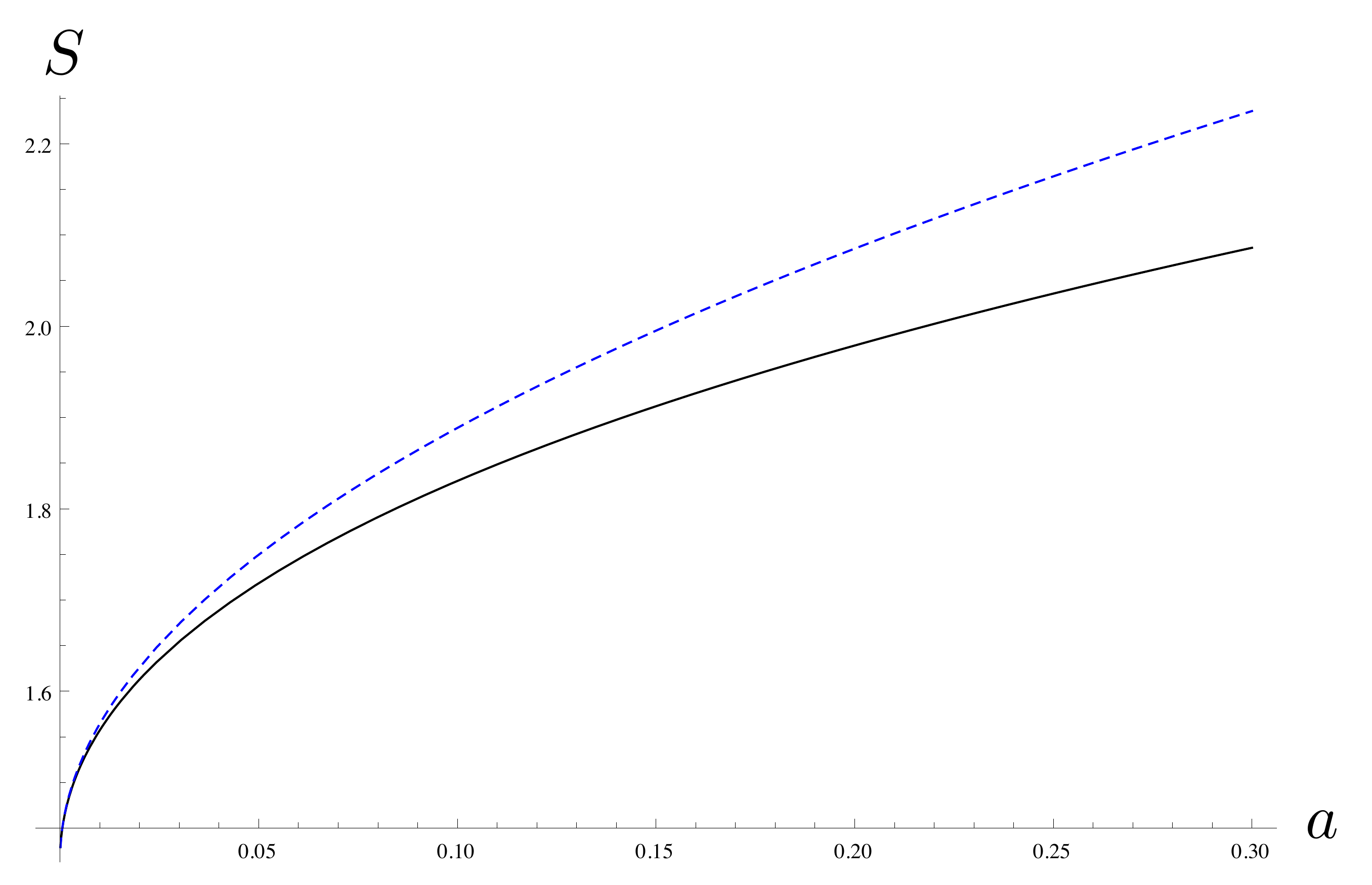}\\
  \caption{Dashed (leading approximation), solid (exact) $(h^0=h_1=h_2=h_3=p^3=1$).}\label{sapplot}}
\end{figure}

The second example is based on the prepotential
\begin{equation}
F(Y) = - \left(Y^1 Y^2 Y^3 + a \, (Y^3)^3 \right)/Y^0  = i \left(Y^0\right)^2 \left(STU + a \, U^3 \right)\;,
\label{t-sugra}
\end{equation}
where $S= - i Y^1/Y^0, T = - i Y^2/Y^0, U=-i Y^3/Y^0$ and $a> 0$.
When embedded into heterotic string theory, the $U^3$-term constitutes a perturbative (one-loop)
correction, and the prepotential  \eqref{t-sugra} describes the perturbative chamber $S >> T > U$ \cite{Harvey:1995fq}.
We consider solutions that are supported by charges $(q_0, p^3)$ and
fluxes $(h_1, h_2, h_3, h^0)$.  We demand that these satisfy the Hamiltonian constraint \eqref{hamil}, 
$q_0 h^0 = p^3 h_3$, and we take the fluxes to be all positive.
We seek a supersymmetric solution, and hence we proceed to solve
\eqref{Y-ansatz}, where we set $e^{i \delta} =i$, for concreteness.
These equations constitute equations for $S, T, U$ and $v_2$ and 
they take the form (we set $g=1$)
\begin{eqnarray}
q_0 + p^3 \left(ST + 3 a \, U^2 \right) &=& 2 v_2 \, h^0 \left( STU + a \, U^3 \right) \;, \nonumber\\
p^3 T &=& v_2 \, \left(h_1 + h^0 \, TU \right) \;, \nonumber\\
p^3 S &=& v_2 \, \left(h_2 + h^0 SU \right) \;, \nonumber\\
6 a \, p^3 U &=& v_2 \, \left(h_3 + 3 a h^0 U^2 + h^0 ST \right) \;.
\label{eq:stuv2-eq}
\end{eqnarray}
We focus on a solution satisfying
\begin{equation}
h_2 \, T = h_1 S \;,
\label{eq:S=T}
\end{equation}
which is consistent with the second and third equations. 
We find that we can numerically
construct an exact solution to \eqref{eq:stuv2-eq}
satisfying \eqref{eq:S=T} 
that has the feature that it only exists for
non-vanishing $a$.  In particular, the field $U$ blows up as $a \rightarrow 0$, with
$v_2$ shrinking to zero in this limit.  This is shown in Figure \ref{BB}.
Thus, this
solution only exists due to quantum corrections: they turn a non-$AdS_2$ geometry
into an $AdS_2$ geometry.  The solution can also be constructed
iteratively, as follows.   We expand $S, T, U, v_2$ as follows,
\begin{eqnarray}\label{expansion}
S&=& s_0+ s_1\sqrt{a}+s_2 a+\dots \;,\nonumber\\
T&=& t_0 + t_1\sqrt{a}+t_2 a+\dots \;, \nonumber\\
U&=&\frac{u_0}{\sqrt{a}}+u_1 +  u_2\sqrt{a}+u_3 a+\dots \;,  \nonumber\\
v_2&=& \alpha_0\sqrt{a}+\alpha_1 a+\dots \;.
\end{eqnarray}
Inserting this ansatz into the attractor combination
$ \Delta_I=\hat Q_I-i \, v_2 \hat h_I$ yields the expansion
\begin{eqnarray}
 \Delta_i &=& \Delta_i^{(0)}+\Delta_i^{(1)}\sqrt{a}+\dots \;, \;  i=0,1,2,  \nonumber\\
  \Delta_3 &=& \Delta_3^{(1)}\sqrt{a}+\dots \;.
  \label{eq:Delta-appox}
\end{eqnarray}
This system can be solved iteratively, order by order in the expansion parameter $a$. 
When doing so, we find $u_1=0$. To simplify the expressions below, we set $u_1=0$ in \eqref{expansion} from the start.
Then, to lowest
order, the system $ \Delta^{(0)}_i=0$ yields
\begin{equation}\label{Ord0}
  s_0 \,  t_0=\frac{h_3+h^0u_0^2}{h^0} \;\;\;,\;\;\;\alpha_0 \, u_0=\frac{p_3}{h^0} \;\;\;,\;\;\; h_2 \, t_0 = h_1 \, s_0
  \;.
\end{equation}
This determines $s_0, t_0$ and $\alpha_0$ in terms of $u_0$ which, at this order, remains undetermined, but
gets determined 
recursively by going
to the next order.  At the next order, the system  $ \Delta_I^{(1)} =0$ determines the values of the parameters
$s_0, t_0, u_0, \alpha_0$ to be 
\begin{equation}
 s_0=	\sqrt{\frac{2h_2h_3}{h^0h_1}}\;\;,\;\; t_0= \sqrt{\frac{2h_1h_3}{h^0h_2}}\;\;,\;\;
u_0= \sqrt{\frac{h_3}{h^0}}\;\;,\;\;\alpha_0  =\frac{p^3}{\sqrt{h^0h_3}} \;.
\label{values:ordz}
\end{equation}
In addition, using $h_2 \, t_1 = h_1 \, s_1$ (which follows from \eqref{eq:S=T}), we obtain
\begin{equation}
	 t_1= \frac32 \frac{h_1}{ \sqrt{h^0 h_3}} \;\;,\;\; \alpha_1=-\sqrt{\frac{h_1h_2}{2h^0 h_3^3}} \, p^3\;.
\end{equation}
The value of $u_2$ is again determined recursively by going to the next order.  The approximate solution, obtained
by solving \eqref{eq:Delta-appox}, can be compared with the exact solution, see
Figures \ref{BB} and \ref{sapplot}.   The values \eqref{expansion} are invariant under uniform scalings of the charges
and the fluxes, since the attractor equations \eqref{Y-ansatz} scale uniformly.

In addition, to determining the values of $S,T,U, v_2$, we also need to determine the values
of $\Upsilon, Y^0$ and $v_1$ (or equivalently of $\Xi$).  We expand these fields as
\begin{eqnarray}
 \sqrt{-\Upsilon} &=& \lambda_0+\lambda_1\sqrt{a}+\dots \;, \nonumber\\
Y^0 &=& y_0 \sqrt{a}+y_1 a+\dots \;, \nonumber\\
v_1 &=& \beta_0 \, \sqrt{a} + \beta_1 \, a + \dots \;.
\label{ansatz-y-ups}
\end{eqnarray}
Inserting this into \eqref{eq:v1} yields the following value for $\beta_0$,  
\begin{equation}
\beta_0= \frac{3}{8 \, \sqrt{h^0 \, h_3^3}} \;.
\label{vv1-stu3}
\end{equation}
On the other hand, inserting the ansatz \eqref{ansatz-y-ups} into
\eqref{eq:U-Y-Ups} and \eqref{value-Ups} leads to a determination of the lowest order coefficients 
$\lambda_0$ and $y_0$. We find (for $h_3, h^0$ and $p^3$ positive, for concreteness) 
\begin{eqnarray}
\lambda_0 &=& \frac{16 (3289 + 592 \sqrt{30})}{1083} h_3 p^3 \;, \nonumber\\
y_0 &=& \frac{37 + 8 \sqrt{30}}{114} \frac{\sqrt{h^0} p^3}{\sqrt{h_3}}\;.
\end{eqnarray}

\begin{figure}[t]
   \center{\includegraphics[scale=0.4]{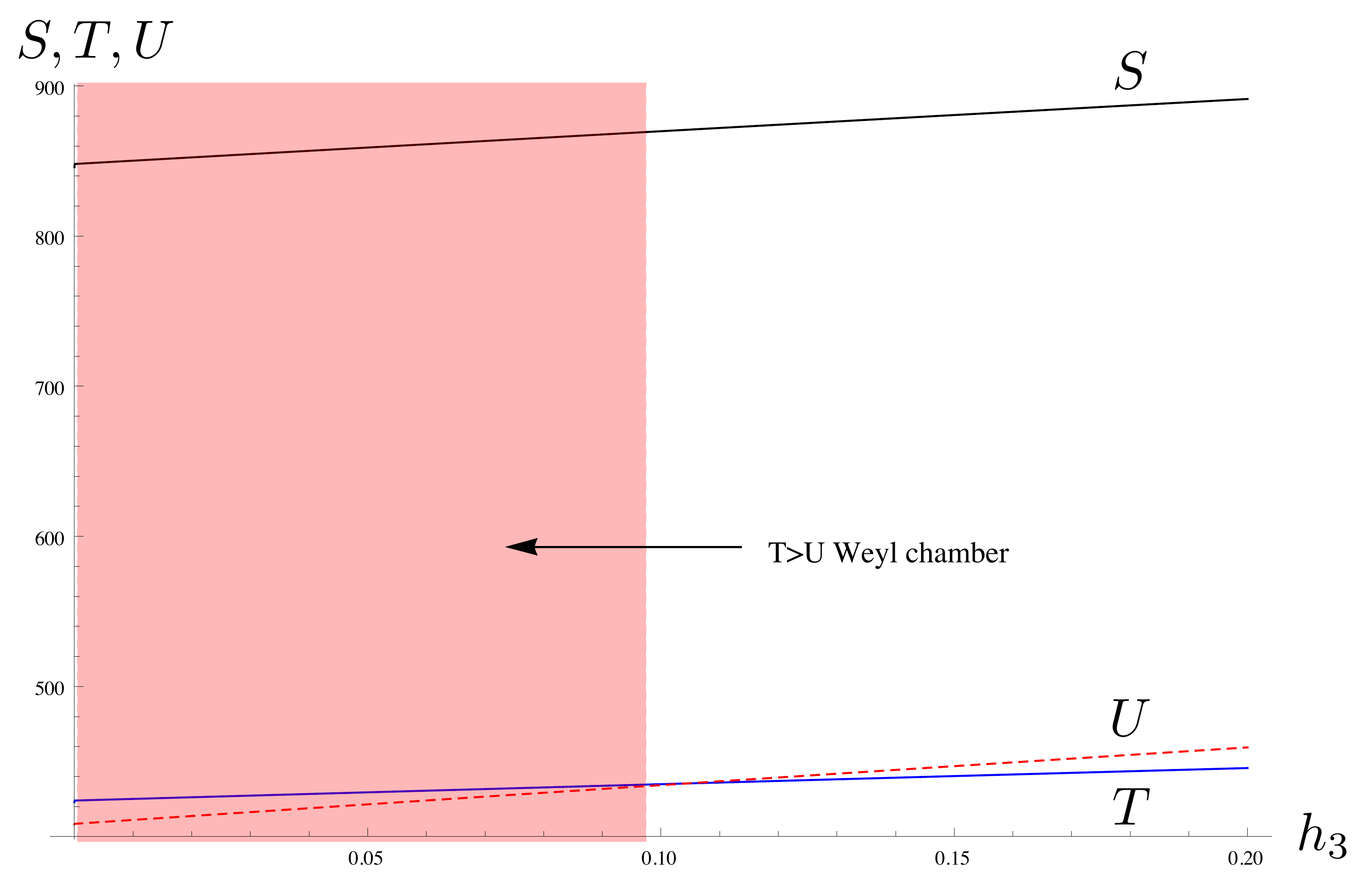}}\\
   \caption{Ensuring $S>>T>U>>1$ ($h^0=0.00001, h_1=1, h_2=2, a=1/3$).}
    \label{STUhier}
\end{figure}

To ensure that the exact solution to \eqref{eq:stuv2-eq} is in the perturbative chamber $S >> T >  U>>1 $ (we set $ a = \tfrac13$) , we have to choose the fluxes
appropriately.  By choosing $h^0$ to be small and taking $h_2/h_1 >1$ we can ensure  $S>>T>>1$.
Picking $h_3$ accordingly, we can then also enforce $T >  U>>1 $, as depicted in Figure \ref{STUhier}.
In addition, the values of the fluxes may be chosen in such a way to ensure that there exists an interpolating
solution that connects the  
$AdS_2 \times \mathbb{R}^2$ background discussed here to a solution that asymptotes to $AdS_4$.  This will be discussed
in section \ref{interpol}. Interestingly, we will obtain a flow to $AdS_4$ that only exists when 
$a \neq 0$.

Finally, we note that we could add a term proportional to $i c \, (Y^{(0)})^2$ to the prepotential
\eqref{t-sugra} and repeat the analysis given above.  
Such a term also represents a perturbative correction in heterotic string theory.
Its presence would lead to a modification of the solution given above.  We have chosen not to include
such a term in our analysis, for simplicity.

\subsection{Variational equations with higher-derivative terms}

Next, we turn to the entropy function \eqref{eq:entropy-total-2} in the presence of higher-derivative terms, 
and we compute the associated extremization
equations for the fields $\Xi, \Upsilon$ and $Y^I$. The quantities $F,K,W$ are now given by \eqref{eq:homo-F}
and \eqref{K-W-full}. 
Although we will be interested in the black brane case ($k=0$), we
keep $k$ as a bookkeeping device. We follow the exposition given in \cite{Cardoso:2006xz}.

Varying with respect to $\Xi$ gives
\begin{eqnarray}
  \label{eq:valueUcorr} 
  && \Sigma + \left( {\cal Q}_I - F_{IK} \, {\cal P}^K\right) 
  N^{IJ} \left({\cal Q}_J -{\bar F}_{JL} \, {\cal P}^L\right)
   - \frac{8 k }{\sqrt{-\Upsilon}} \, K(Y, \bar Y)
     \\
  &&
  - i(F_\Upsilon-\bar F_\Upsilon) \Big [- 4 \Upsilon + 64
   (k^2-\Xi^{-2}) - 16 k \sqrt{-\Upsilon} \Big] - 64 g^2 \,  \Upsilon^{-1} \,
    \left[K(Y, \bar Y)\right]^2 \, V(Y, \bar Y) =0 \;. \nonumber
   \end{eqnarray}
Expressing the combination ${\cal Q}_I - F_{IK} \, {\cal P}^K$ in terms of the combination
\eqref{hatQhat}  gives
\begin{equation}
{\cal Q}_I - F_{IK} \, {\cal P}^K = - \Sigma_I \;,
\end{equation}
where 
\begin{equation}
\Sigma_I = - \left({\hat Q}_I + K_I + 2 i \Upsilon F_{\Upsilon I}\right) \;.
\label{der-sigm}
\end{equation}
Here $\Sigma_I = \partial_I \Sigma$, where $\Sigma$ is given in \eqref{comb-QPSIG}.
For $k=0$, \eqref{eq:valueUcorr} becomes
\begin{eqnarray}
  \label{eq:valueUk0} 
   \Sigma + \Sigma_I
  N^{IJ} \Sigma_{\bar J}
    - i(F_\Upsilon-\bar F_\Upsilon) \Big [- 4 \Upsilon 
   -64 \Xi^{-2}\Big] - 64 g^2 \,  \Upsilon^{-1} \,
    \left[K(Y, \bar Y)\right]^2 \, V(Y, \bar Y) =0 \;. 
\end{eqnarray}

Next, we consider the variation of the entropy function
\eqref{eq:entropy-total-2} with respect to arbitrary variations of the
fields $Y^I$ and $\Upsilon$ and their complex conjugates. Denoting
this variation by $\delta= \delta Y^I\partial/\partial Y^I+ \delta\bar
Y^I\partial/\partial \bar Y^I+ \delta\Upsilon\partial/\partial
\Upsilon+ \delta\bar \Upsilon\partial/\partial \bar\Upsilon$, we
obtain
\begin{eqnarray}
  \label{eq:deltaentro}  
\tfrac12   \delta {\cal E} &=& \Xi \left[
   \mathcal{P}^I \,\delta(F_I+\bar F_I) -  \mathcal{Q}_I \,\delta(Y^I+\bar
  Y^I) \right] \nonumber\\
  &&  
  +\tfrac1{2} i \, \Xi \left[({\cal Q}_K -\bar F_{KM} \,{\cal P}^M)
  N^{KI} \,\delta F_{IJ} \, N^{JL} ( {\cal Q}_L - {\bar F}_{LN}
  \,{\cal P}^N )   - \mathrm{h.c.} \right] 
  \nonumber \\
  && 
  - 4 i (- \Upsilon)^{-1/2}\, (k \Xi-1) \,
  \left[ (F_I - {\bar F}_I) \,\delta(Y^I + {\bar Y}^I) 
    - (Y^I - {\bar Y}^I) \, \delta (F_I + {\bar F}_I) \right]
  \nonumber\\
  &&
  + i \left[ 2 \Xi \, \Upsilon - 32 (k^2 \Xi + \Xi^{-1} -2k) 
    +16 \sqrt{-\Upsilon} \right] 
  \delta (F_{\Upsilon} - {\bar F}_{\Upsilon}) \nonumber\\
  && 
  + i \, \Xi  \left[\delta\Upsilon\,F_{\Upsilon I} N^{IJ} ( {\cal
   Q}_J - {\bar F}_{JL} \,{\cal P}^L) - \mathrm{h.c.} \right] 
   \nonumber\\
   &&- 2 i (-\Upsilon)^{-3/2} \, (k \Xi-1) \, 
   ({\bar Y}^I F_I - Y^I {\bar F}_I )
   \, \delta \Upsilon \nonumber\\
   &&+  i  (F_{\Upsilon} - {\bar F}_{\Upsilon}) \left[ \Xi - 
     4(-\Upsilon)^{-1/2} \, (1 + k \Xi) \right] \, \delta \Upsilon \nonumber\\
     && - 32 g^2 \Xi \, \Upsilon^{-1} \left[ 2 K \, V \, \delta K  - \Upsilon^{-1} \, K^2 \,  V \, \delta \Upsilon
     + K^2 \, \delta V \right] \;,
\end{eqnarray}
where we took into account that the variable $\Upsilon$ is real. For $k=1$ this reduces to the expression derived in
\cite{Cardoso:2006xz}.

Restricting to variations $\delta Y^I$ gives
\begin{eqnarray}
  \label{eq:variational-y}
  &&
  \Xi \left(\mathcal{Q}_I - F_{IJ} \, \mathcal{P}^J\right) 
   -\tfrac1{2} i \, \Xi \left({\cal Q}_K -\bar F_{KM} \,{\cal P}^M
  \right) 
  N^{KP} \,F_{PIQ} \, N^{QL} \left({\cal Q}_L -{\bar F}_{LN}\,{\cal
  P}^N \right) \nonumber\\
  && 
  + 4 i (- \Upsilon)^{-1/2} (k \Xi-1) 
  \left[F_I - {\bar F}_I - F_{IJ} (Y^J -{\bar Y}^J) \right]
  \\
  &&
  -  i \left[ 2 \Xi \, \Upsilon - 32 (k^2 \Xi + \Xi^{-1} -2k) 
    +16 \sqrt{-\Upsilon} \right] 
  F_{\Upsilon I} 
  + 32 g^2 \Xi \, \Upsilon^{-1} \left[ 2 K \, K_I \, V 
     + K^2 \, \, V_I \right]   = 0 \;. \nonumber
\end{eqnarray}
Using $\hat{h}_I = \bar{\hat{h}}_I - i N_{IJ} \, h^J$, we obtain for $V_I$,
\begin{equation}
V_I = i N^{KP} \, F_{PQI} \, N^{QL} \, {\bar{\hat h}}_K \, {\bar{\hat h }}_L 
+ 2 K^{-2} \left[ |W|^2 \, K_I + {\bar W} \, K\, \hat{h}_I \right] \;.
\end{equation}
Focussing on the black brane case ($k=0$), we obtain from \eqref{eq:variational-y},
\begin{eqnarray}
   &&
  - \Xi \, \Sigma_I
   -\tfrac1{2} i \, \Xi \, {\bar \Sigma}_{\bar K}
  N^{KP} \,F_{PIQ} \, N^{QL} \, {\bar \Sigma}_{\bar L} \nonumber\\
  && 
  - 4 (- \Upsilon)^{-1/2} 
  \left[K_I + 2 i \Upsilon F_{\Upsilon I} \right]
  \\
  &&
  -  i \left[ 2 \Xi \, \Upsilon - 32  \Xi^{-1}  
    +16 \sqrt{-\Upsilon} \right] 
  F_{\Upsilon I} 
  + 32 g^2 \Xi \, \Upsilon^{-1} \left[ 2 K \, K_I \, V 
     + K^2 \, \, V_I \right]   = 0 \;, \nonumber
\end{eqnarray}
and, using \eqref{der-sigm}, we get
\begin{eqnarray}
\label{attrac-Y}
   &&
   \Xi \, \left( \hat{Q}_I  + 64 g^2 \Upsilon^{-1} K {\bar W} \hat{h}_I \right)
   -\tfrac1{2} i \, \Xi\, 
  N^{KP} \,F_{PIQ} \, N^{QL} \left( {\Sigma}_{\bar K}
 {\Sigma}_{\bar L} - 64 g^2 \, \Upsilon^{-1} \, K^2 \,  {\bar{\hat h}}_K \, {\bar{\hat h }}_L 
 \right)
   \nonumber\\
  && 
    + \Xi K_I \left(
 1 - 4 (- \Upsilon)^{-1/2} \, \Xi^{-1} +  64 g^2  \, \Upsilon^{-1} 
  K  \, V + 64 g^2  \, \Upsilon^{-1} \,|W|^2 
\right) \nonumber\\
&&  +  i \left[  32\,  \Xi^{-1}  
    -8 \sqrt{-\Upsilon} \right] 
  F_{\Upsilon I} 
   = 0 \;.
\end{eqnarray}
Observe that all the terms transform as vectors under symplectic transformations.

Next, let us restrict \eqref{eq:deltaentro}  to variations $\delta \Upsilon$ (recall that $\Upsilon$ is a real variable).  Setting $k=0$, and using
\begin{eqnarray}
K_{\Upsilon} &=& i \left( {\bar Y}^I F_{\Upsilon I} - Y^I {\bar F}_{\Upsilon {\bar I}}
\right) \;, \\
V_{\Upsilon} &=& i \left( N^{KP} \, F_{PQ\Upsilon} \, N^{QL} \, {\bar{\hat h}}_K \, {\bar{\hat h }}_L 
- {\rm h.c.} \right)
+ 2 K^{-2} \left[ |W|^2 \, K_{\Upsilon} - K \left( {\bar W} \, F_{\Upsilon I}  
+ W \, {\bar F}_{\Upsilon \bar I} \right) h^I \right] \;, \nonumber
\end{eqnarray}
we obtain
\begin{eqnarray}
\label{var-ups-k0}
   &&
   \Xi \, \left[ i F_{\Upsilon I} N^{IJ} \left( - \Sigma_J - 64 i  g^2 \Upsilon^{-1} K {\bar W} \,
   N_{JK} h^K \right) + {\rm h.c.} \right] \nonumber\\
   &&   +\tfrac1{2} \, i \, \Xi \, \left[ 
  N^{KP} \,F_{\Upsilon PQ} \, N^{QL} \,   
  \left( {\Sigma}_{\bar K}
 {\Sigma}_{\bar L} - 64 g^2 \, \Upsilon^{-1} \, K^2 \,  {\bar{\hat h}}_K \, {\bar{\hat h }}_L 
 \right) - {\rm h.c.} \right]
  \nonumber\\
  && + 2\, \Xi \left( 16 g^2 \Upsilon^{-2} K^2 V +  (- \Upsilon)^{-3/2} \Xi^{-1} K \right) \nonumber\\
  && 
    - \Xi K_{\Upsilon} \left(
 - 4 (- \Upsilon)^{-1/2} \, \Xi^{-1} +  64 g^2  \, \Upsilon^{-1} 
  K  \, V + 64 g^2  \, \Upsilon^{-1} \,|W|^2 
\right) \nonumber\\
  && 
  + i \left(\Xi - 4  (- \Upsilon)^{-1/2} \right) \left(F_{\Upsilon} - \bar{F}_{\Upsilon} \right)
    \\
  &&
  + i \left[ 2 \Xi \, \Upsilon - 32  \Xi^{-1}  
    +8 \sqrt{-\Upsilon} \right] 
  \left( F_{\Upsilon \Upsilon} -  {\bar F}_{\Upsilon \Upsilon} \right)
        = 0 \;. \nonumber
\end{eqnarray}
Observe that not all combinations are symplectic functions.  This is so, because the derivative
$\partial/\partial
\Upsilon$, when acting on a symplectic function, does not yield a symplectic
function \cite{deWit:1996ix}.
To obtain combinations that are symplectic functions, we may use the 
mixed derivative $Y^I\partial/\partial Y^I+
\bar Y^I\partial/\partial \bar Y^I+ 2 \Upsilon\partial/\partial
\Upsilon$, where $\Upsilon$ is real so that
$\partial/\partial\Upsilon$ acts on both $\Upsilon$ and $\bar
\Upsilon$ \cite{Cardoso:2006xz}.
Then, using the homogeneity relation
\begin{equation}
Y^I K_I + {\bar Y}^I K_{\bar I} + 2 \Upsilon K_{\Upsilon} = 2 K \;,
\end{equation}
we obtain
\begin{eqnarray}
\label{attrac-Ups}
&& - 2 \Xi K \left(
 1 - 4 (- \Upsilon)^{-1/2} \, \Xi^{-1} +  64 g^2  \, \Upsilon^{-1} 
  K  \, V + 64 g^2  \, \Upsilon^{-1} \,|W|^2 
\right) \nonumber\\
 && + 2 \Xi \left( 16 g^2 \Upsilon^{-2} K^2 V +  (- \Upsilon)^{-3/2} \Xi^{-1} K \right) \nonumber\\
  &&+ 2 i \left(\Xi - 4  (- \Upsilon)^{-1/2} \right) \left(F_{\Upsilon} - \bar{F}_{\Upsilon} \right)
    \nonumber\\
    && + \Xi \left[ Z(Y) +  64 g^2  \, \Upsilon^{-1} \,K \, |W|^2 - 2 i  \Upsilon F_{\Upsilon I} 
    N^{IJ} \Sigma_{\bar J} + {\rm h.c.} \right] =0 \;,
\end{eqnarray}
where $Z(Y)$ denotes the extension of \eqref{K-Z} given by
\begin{equation}
Z(Y) =  p^I F_I(Y, \Upsilon) - q_I Y^I \;.
\end{equation}
Observe that each line of \eqref{attrac-Ups} constitutes a symplectic function.

Inserting \eqref{eq:valueUk0} into the entropy function \eqref{eq:entropy-total-2} with $k=0$ gives 
\begin{equation}
{\cal E} =  \frac{8 K}{\sqrt{-\Upsilon}} + 16 i (F_{\Upsilon} - {\bar F}_{\bar \Upsilon} )
( \sqrt{-\Upsilon}  - 8 \Xi^{-1} )  \;.
\label{entro-densit}
\end{equation}
The black brane entropy density is given by \eqref{entro-densit}, with $Y^I, \Upsilon$ and $\Xi$ 
expressed in terms of charges and fluxes by 
solving the extremization equations \eqref{attrac-Y}, \eqref{attrac-Ups} and \eqref{eq:valueUk0}.  
To solve these equations, one may proceed iteratively by power expanding in $\Upsilon$.

\subsubsection{An example}

In the presence of higher-derivative interactions, the extremization equations \eqref{attrac-Y}, \eqref{attrac-Ups} and \eqref{eq:valueUk0} take a rather complicated form.  One way to deal with these complications is to
expand $F(Y, \Upsilon)$ in powers of $\Upsilon$, 
\begin{equation}
F(Y, \Upsilon) = \sum_{g=0}^{\infty} \Upsilon^g \, F^{(g)}(Y) \;,
\end{equation}
and to solve the extremization equations order by order in $\Upsilon$.

In the following, we will focus on a particular model with an $F^{(0)}$ and $F^{(1)}$ only,
namely
\begin{equation}
F(Y, \Upsilon) = - \frac{Y^1 Y^2 Y^3}{Y^0} - \Upsilon \, c_1 \, \frac{Y^1}{Y^0} \;,
\label{eq:heteroticprep}
\end{equation}
with $c_1 > 0$.  In ungauged supergravity, this model captures features
of $N=4$ models in the presence of $R^2$ interactions.  In particular,  it allows for supersymmetric 
small black holes, which
are solutions that only exist due to the presence of  the term proportional to $c_1$ in \eqref{eq:heteroticprep} \cite{Dabholkar:2004yr,Dabholkar:2004dq}.  Consider a small black hole that 
carries charges
 $(q_0, p^1)$ with $q_0 p^1 < 0$.
 The supersymmetric attractor equations for the $Y^I$ are $Y^I - {\bar Y}^I = i p^I$
 and $F_I - {\bar F}_I = i q_I$.  They can be readily solved 
 for the model \eqref{eq:heteroticprep} \cite{Sen:2004dp}. The attractor values for $T = -i Y^2/Y^0$ and 
 $U=-i Y^3/Y^0$ are zero,  while $S= -i Y^1/Y^0$ and $Y^0$ take non-vanishing values that exhibit the
 following scaling behavior with $c_1$ \cite{LopesCardoso:2004xf},
 \begin{eqnarray}
 S+ \bar S &=& \frac{s_0}{\sqrt{c_1}} \;, \nonumber\\
 Y^0 = \bar Y^0 &=& y_0 \, \sqrt{c_1} \;,
 \label{small-sol}
 \end{eqnarray}
 where $s_0$ and $y_0$ are given by $s_0 =  \sqrt{|q_0 p^1/\Upsilon|} \; , \; 
 y^0 = \sqrt{| \Upsilon \, p^1/q_0|}$, and the field
 $\Upsilon$ takes the value $\Upsilon = - 64$ at the horizon \cite{LopesCardoso:1998wt}.  The entropy ${\cal E}$,
 which is non-vanishing, 
 is determined in terms of $S + \bar S$ as 
 ${\cal E} = 32 \pi \sqrt{|q_0 p^1|} \,   \sqrt{c_1}$.
 
Now consider turning on fluxes $(h^0, h_1, h_2, h_3)$. 
For  large charges, and for a certain range of fluxes, we expect that there exist black brane solutions whose
near-horizon geometry can be approximated by the geometry of a small black hole.
  Thus, we expect to be able to construct black brane solutions
 to the extremization equations
 \eqref{attrac-Y}, \eqref{attrac-Ups} and \eqref{eq:valueUk0}
  that are supported by charges $(q_0, p^1)$ and have the scaling behavior
 \begin{eqnarray}\label{ansatz-scaling}
S+ \bar S&=&  \frac{s_0}{\sqrt{c_1}} + s_1 +\dots  \;, \nonumber\\
T + \bar T&=& t_0 + t_1\sqrt{c_1} +\dots \;, \nonumber\\
U + \bar U&=& u_0 + u_1 \sqrt{c_1} +\dots \;, \nonumber\\
Y^0 &=& y_0 \sqrt{c_1} + y_1 c_1 + \dots \;, \nonumber\\
\sqrt{-\Upsilon} &=& \lambda_0 + \lambda_1 \sqrt{c_1} + \dots \;, \nonumber\\
\Xi &=&  \xi_0 + \xi_1 \sqrt{c_1}+  \dots \;.
\end{eqnarray}
We note the ansatz \eqref{ansatz-scaling} for the scaling behavior is analogous to
the one discussed in \eqref{expansion} for the model \eqref{t-sugra}.  

We proceed to solve the attractor equations \eqref{attrac-Y}, \eqref{attrac-Ups} and \eqref{eq:valueUk0} iteratively by power expanding in $c_1$, in a manner similar to 
what we did for the model \eqref{t-sugra}. 
Concretely we found a solution for $h^0 = 0 = h_1$ and with $q_0<0$ and $p^1>0$.\footnote{Given 
that we turned off the fluxes $h^0$ and $h_1$ it is likely impossible to extend this case to an 
asymptotic $AdS_4$ solution. Finding such an interpolating solution with higher derivative 
corrections is clearly outside the scope of this paper.} Plugging the ansatz 
\eqref{ansatz-scaling} into the five attractor equations  \eqref{attrac-Y}, \eqref{attrac-Ups} and \eqref{eq:valueUk0} 
and expanding each one in a power series in $c_1$, the five leading order equations are solved by 
\begin{equation}\label{leading_solution_c1}
s_0 = 0.441 \frac{\sqrt{|q_0| p_1}}{\lambda_0} \quad , \quad t_0 = u_0 = 1.306 \sqrt{\frac{|q_0|}{p_1}} \quad , \quad y_0 = 1.190 \frac{\sqrt{p_1}\lambda_0}{\sqrt{|q_0|}} \quad , \quad \xi_0 = \frac{4.898}{\lambda_0}\ ,
\end{equation}
whereas $\lambda_0$ is not constrained at this order. Inserting \eqref{leading_solution_c1} into the equations obtained by expanding each of the attractor equations to their next order in $c_1$ gives five constraints involving $\lambda_0$ and the subleading coefficients of the expansion \eqref{ansatz-scaling}. However, $\lambda_0$ and $\lambda_1$ are again not constrained and we find
\begin{eqnarray}\label{subleading_solution_c1}
s_1 &=& -0.441 \frac{\sqrt{|q_0| p_1}\lambda_1}{\lambda_0^2} = -s_0 \frac{\lambda_1}{\lambda_0} \ , \nonumber \\
 t_1 &=& u_1 = 0 \ , \nonumber  \\
y_1 &=& 1.190\frac{\sqrt{p_1}\lambda_1}{\sqrt{|q_0|}} = y_0 \frac{\lambda_1}{\lambda_0} \ ,  \nonumber \\ 
\xi_1 &=& -\frac{4.898 \lambda_1}{\lambda_0^2} = -\xi_0  \frac{\lambda_1}{\lambda_0}\ .
\end{eqnarray}

Observe that whereas the small black hole
solution \eqref{small-sol} represents an exact solution to the supersymmetric attractor
equations of ungauged supergravity, the black brane solution discussed here will receive
corrections order by order in $c_1$. This is due to the complicated form of the attractor
equations \eqref{attrac-Y}, \eqref{attrac-Ups} and \eqref{eq:valueUk0}. Moreover, naively 
it appears as if the solution we found does not depend on the values of the fluxes $h_2$ and $h_3$. However,
this is an artefact of our truncation to the lowest orders in the $c_1$-expansion. The full attractor equations
do depend on $h_2$ and $h_3$ and we expect the more subleading coefficients in the expansion \eqref{ansatz-scaling}
to depend on them as well. It would be interesting to pursue this point further.

\section{Interpolating solutions}
\label{interpol}

In the following, we will switch off higher-derivative interactions and 
consider interpolating extremal black brane solutions in the presence of quantum corrections to the prepotential.  We will focus
on solutions that only exist due to the presence of these quantum corrections.
For concreteness, we pick the STU-model described by \eqref{t-sugra}, and we construct
solutions
that interpolate between a near-horizon geometry $AdS_2 \times \mathbb{R}^2$ and 
an $AdS_4$ geometry. These solutions will be supported by fluxes $(h^0, h_1, h_2, h_3)$ as well as by charges
$(q_0, p^1, p^2, p^3)$.  More precisely, we will consider what happens when one or two of the magnetic charges
$p^A$ ($A=1,2,3$) are turned off.  The fluxes and the charges are 
subjected to the Hamiltonian constraint \eqref{hamil}.  

The choice of the fluxes ensures that the flux potential \eqref{flux-pot} has $AdS_4$ extrema.
For the prepotential \eqref{t-sugra} we find various such extrema.  We will focus on two of them,
as follows.  We take the fluxes $(h^0, h_1, h_2, h_3)$ to be all positive.
The first extremum is of standard type, i.e. it occurs for the uncorrected prepotential ($a=0$).
The values of the scalar fields $S, T, U$ at this extremum 
are given by
\begin{equation}
   S= \sqrt{\frac{h_2 h_3}{h^0h_1}}\;\;\;,\;\;\;\;  T = \sqrt{\frac{h_1 h_3}{h^0h_2}}
\;\;\;,\;\;\;\;     U =\sqrt{\frac{h_1 h_2}{h^0h_3}} \;.
\label{ads-I}
\end{equation}
These values will get corrected when switching on $a$.
At the extremum \eqref{ads-I} the flux potential takes the value 
\begin{equation}
 V_{F} = -6\sqrt{h^0h_1h_2h_3} \;.
\end{equation}
In the figures given below, this extremum is denoted by type 1 $AdS_4$ fixed point and is represented by a blue dot.

The second extremum is not of standard type, and only exists in the presence of quantum corrections, i.e.
when $a \neq 0$.  The values of the scalar fields $S, T, U$ at this extremum 
are, to leading order in $a$, given by
\begin{equation}
   S= h_2\left(\frac{3}{h^0h_3}\right)^{1/2} \, \sqrt{a} \;\;\;,\;\;\;\;  T = h_1\left(\frac{3}{h^0h_3}\right)^{1/2} \,
   \sqrt{a}
\;\;\;\;,\;\;\;     U = \left(\frac{h_3}{3h^0}\right)^{1/2} \, \frac{1}{\sqrt{a}} \;,
\label{ads-II}
\end{equation}
and the value of 
the flux potential at this extremum is, to leading order in $a$, 
\begin{equation}
 V_F =  - 2 \left( \frac{h^0 h_3^3}{3} \right)^{1/2} \, \frac{1}{\sqrt{a}} \;.
\end{equation}
In the figures given below, this extremum is denoted by type 2 $AdS_4$ fixed point and is represented by a red dot.

Next, we construct interpolating black brane solutions that asymptotically flow to one
of these two $AdS_4$ extrema.  These interpolating solutions 
are obtained as solutions to first-order flow equations 
\cite{Cacciatori:2009iz,Dall'Agata:2010gj,Hristov:2010ri,Barisch:2011ui}.\footnote{See \cite{BarischDick:2012gj}
for a discussion of first-order flow equations for extremal black branes in five dimensions.}
They are described by a static line element of the form
\begin{equation}\label{metric ansatz}
 ds^2= -e^{2U}dt^2+e^{-2U} dr^2+ e^{2A}\left(dx^2+dy^2\right) \;,
\end{equation}
where $U=U(r)$ and $A = A(r)$.  The solutions are supported by scalar fields $X^I$.  It is convenient
to introduce rescaled scalar fields $Y^I$ given by\footnote{Note that the field $Y^I$ differs from 
the one introduced in \eqref{eq:rescaledX}.} \cite{Barisch:2011ui}
\begin{equation}
  Y^I = e^A\bar \varphi X^I \;,
\end{equation} 
where the field $\varphi$ denotes a $U(1)$ compensator. The first-order flow equations can then be expressed in
terms of the scalars $Y^I = Y^I (r)$ as follows, 
\begin{eqnarray}\label{first order system}
      \left(Y^I\right) '&=& e^{-\psi-i\gamma} N^{IK}\left(\bar{\hat Q}_K+ige^{2A}\bar{\hat h}_K\right) \;,\nonumber\\
       \psi '&=& 2g e^{-\psi} \Im\left[e^{i\gamma}W(Y)\right] \;, 
 \end{eqnarray}
where $\psi=A+U$, and with $\hat{Q}, \hat{h}$ and $W(Y)$ as defined in \eqref{hatQhat}, \eqref{def-vq} and 
\eqref{K-W-full} (with $F$ restricted to $F(Y)$).
The quantity $e^{2A}$ is determined in terms of the $Y^I$ by
\begin{equation}\label{D-Gauge Y}
   e^{2A} = K(Y, \bar Y) \;, 
\end{equation}
whereas the phase $\gamma$ satisfies
\begin{equation}
e^{ - 2 i \gamma} = \frac{Z(Y) - i g e^{2A} \, W(Y)}{{\bar Z}(\bar Y) + i g e^{2A} {\bar W}( \bar Y)} \;,
\end{equation}
with $Z(Y)$ given in \eqref{K-Z}. 

The first-order flow equations \eqref{first order system} may have fixed points determined by 
\eqref{Y-ansatz}, where $e^{2A} = v_2$ and $e^{i \delta} =  i$.  
One such fixed point was already obtained in 
\eqref{values:ordz} and \eqref{vv1-stu3}, and it arises when the two magnetic charges $p^1,p^2$ are switched off.
Another fixed point occurs when switching off the magnetic charge $p^3$.  In this case the attractor values for
$S, T, U$ and $e^{2A}$
are, to leading order in $a$, given by (we set $g=1$ in the following)
\begin{equation}
   S = \sqrt{\frac{h_3 p^1}{h^0p^2}}
\;\;\;,\;\;\;     T = \sqrt{\frac{h_3 p^2}{h^0p^1}}
\;\;\;\;\;\;       U = \left(\frac{h_3}{h^0p^1 p^2}\right)^{1/6}\left(\frac{q_0}{a}\right)^{1/3} \;\;\;,\;\;\;
e^{2A} = \sqrt{\frac{p^1p^2}{h^0 h_3}} \;,
\end{equation}
while the value of the flux potential at the attractor is, to leading order in $a$, given by
\begin{equation}
V_F\sim -h^0h_3\left(\frac{h_3}{h^0p^1p^2}\right)^{1/6} \left(\frac{q_0}{a}\right)^{1/3} \;.
\end{equation}
Yet another fixed point is obtained when $p^2$ is switched off.  In this case, and taking into account that 
the fluxes $(h^0, h_1, h_2, h_3)$ are all positive, we find the following attractor values
at leading
order in $a$, 
\begin{equation}
   S=\frac{s_0}{\sqrt{a}}\;\;\;,\;\;\;\; T=t_0\sqrt{a}\;\;\;,\;\;\;\; U= \frac{u_0}{\sqrt{a}}\;\;\;,\;\;\;\;e^{2A}=\alpha_0\sqrt{a} \;,
\end{equation}
where the values $s_0$, $t_0$, $u_0$, $\alpha_0$ are rather complicated expressions in terms of charges and fluxes, which
we do not give here.  We only note the relations
\begin{equation}
  p^1=-h_1\left(\frac{s_0\alpha_0}{t_0u_0}\right)<0\;\;,\;\;\;  p^3 = \alpha_0 
  \left(\frac{h_1+h^0t_0u_0}{t_0}\right) >0 \;,
\end{equation}
which constrain the signs of the magnetic charges $p^1, p^3$.  The value of the flux potential at the attractor is,
to leading order, given by
\begin{equation}
 V_F =  - \frac{v(h,p,q)}{\sqrt{a}} \;,
\end{equation}
with $v>0$.

The three fixed points discussed above give rise to $AdS^2 \times \mathbb{R}^2$ geometries that only exist 
due to the presence of the $a$-term in the prepotential \eqref{t-sugra}.  The fixed points with either non-vanishing $p^3$ or non-vanishing $p^1$ and $p^3$ give rise to 
$AdS_2 \times \mathbb{R}^2$ geometries \eqref{line-k} with $v_1, v_2 \sim a^{1/2}$, as can be 
seen using \eqref{eq:v1}. The fixed point with non-vanishing charges $p^1$ and $p^2$ has $v_1 \sim a^{1/3}$ 
and $v^2 = {\cal O} (a^0)$.  All these geometries have in common that in the limit $a \rightarrow 0$, the $AdS_2$
factor $v_1$ shrinks to zero. The two fixed points for which, in addition, also $v_2 \rightarrow 0$, 
have entropy densities that exhibit Nernst behavior in the limit $a =0$.
Note that the ratio $v_1/v_2$ remains finite in this limit. 

\begin{figure}
    \centering
    \includegraphics[scale=0.45]{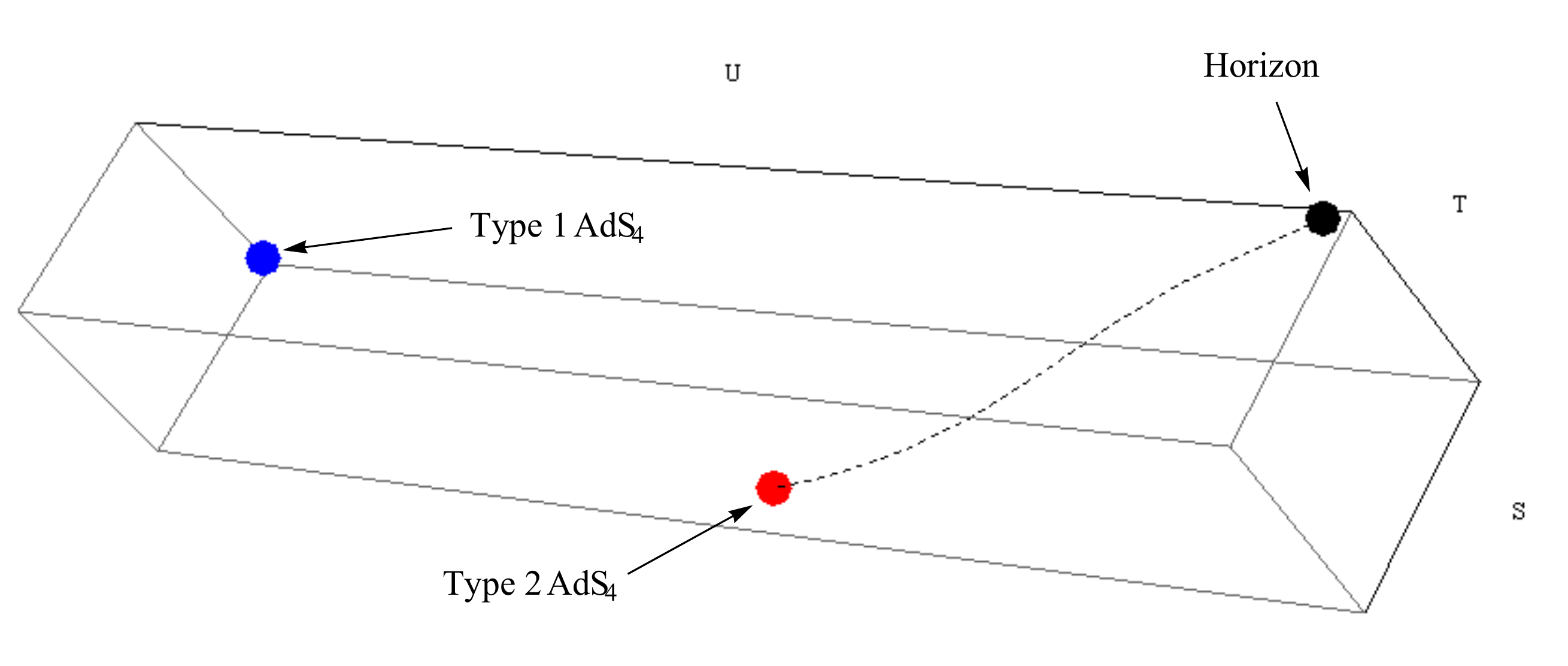}
  \caption{Flow from $AdS_2$ (black dot) to the type 2 $AdS_4$ fixed point  \eqref{ads-II} (red dot)
  $(q_0 =10, p^1 = p^2 =0, p^3=10; h^0=h_1=h_2=h_3 =1).$\label{p3}}
\end{figure}

\begin{figure}
    \centering
    \includegraphics[scale=0.45]{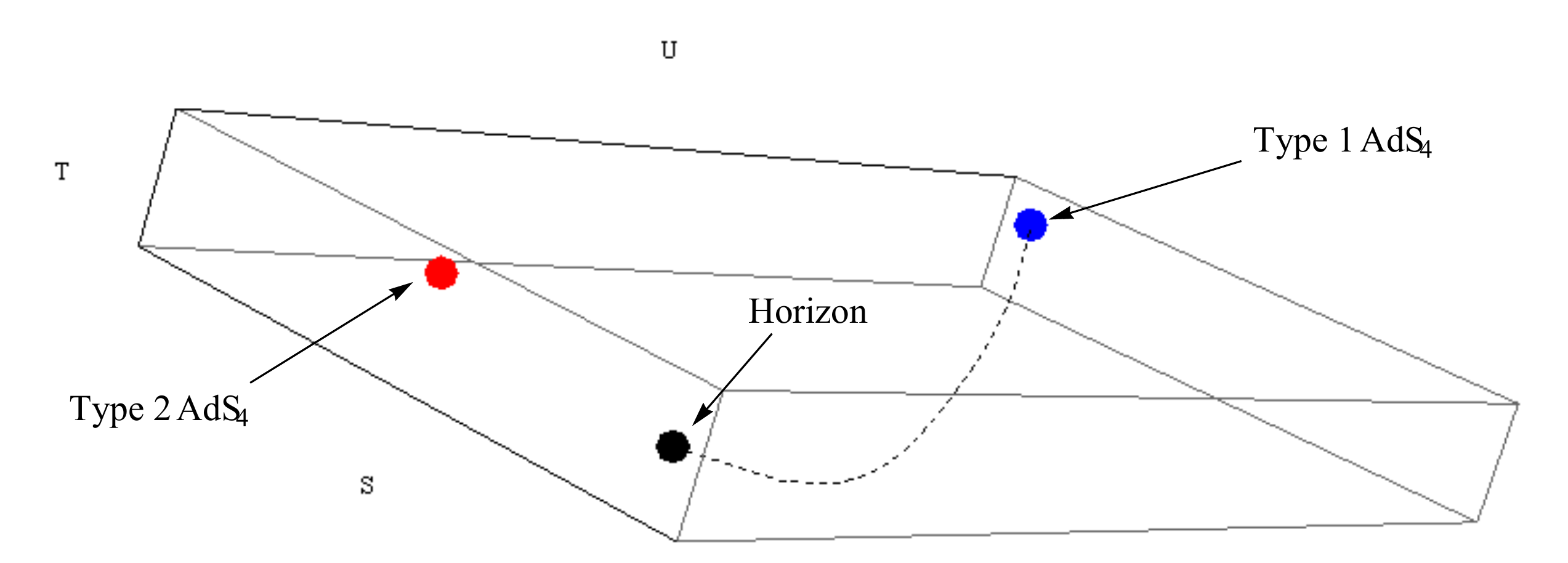}
  \caption{Flow from $AdS_2$ (black dot) to the type 1 $AdS_4$ fixed point  \eqref{ads-I} (blue dot)
  $(q_0 =14, p^1=12, p^2=2, p^3=0; h^0=h_1=h_2=h_3 =1).$ \label{p1p2}}
\end{figure}

\begin{figure}
    \centering
    \includegraphics[scale=0.45]{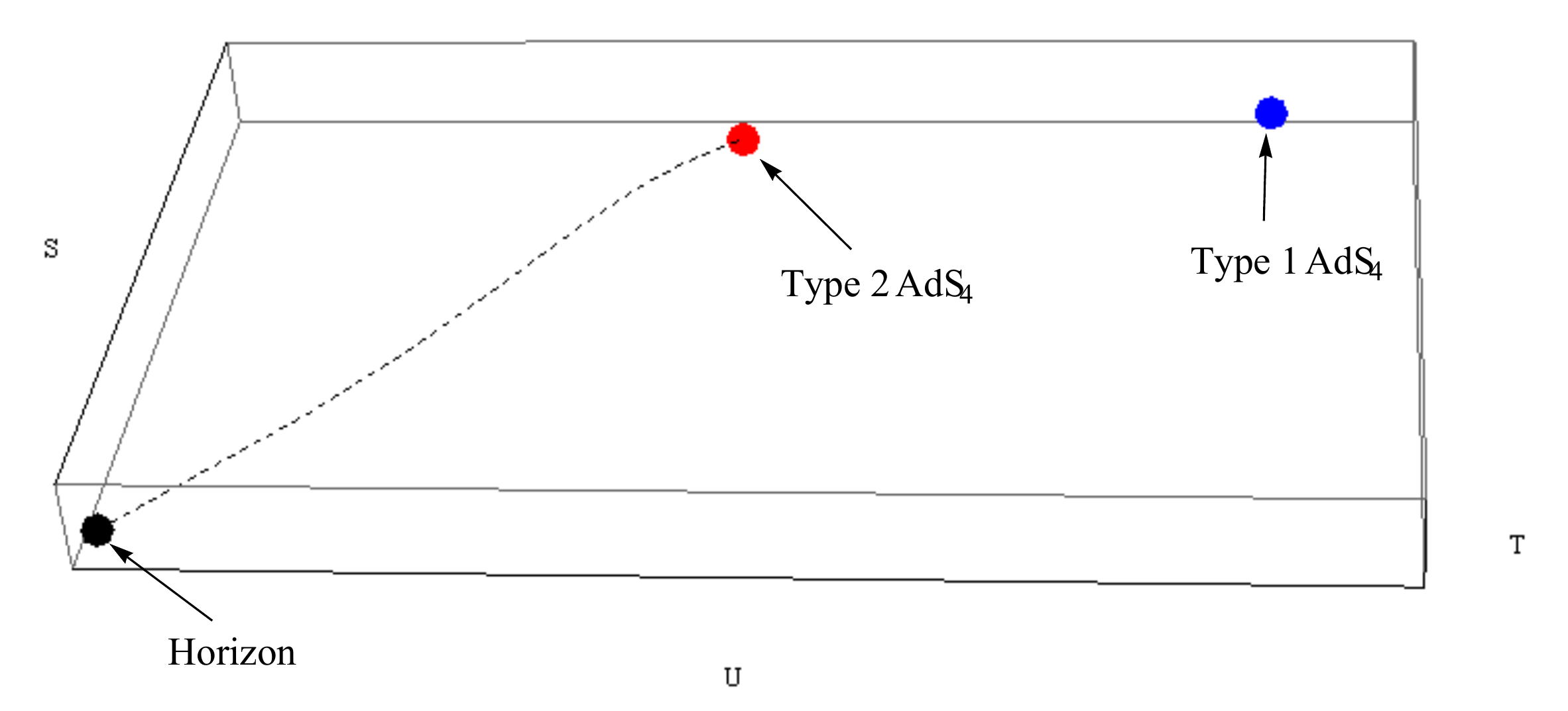}
  \caption{Flow from $AdS_2$ (black dot) to the type 2 $AdS_4$ fixed point  \eqref{ads-II} (red dot)
  $(q_0 =9, p^1=-1, p^2 =0, p^3=10; h^0=h_1=h_2=h_3 =1).$
\label{p2}  }
\end{figure}

Next, we would like to check whether the three $AdS_2$ fixed points can be connected to the two $AdS_4$
fixed points discussed earlier.  This can be done by numerically solving the first-order flow equations
\eqref{first order system}, as explained in appendix \ref{numer}.
Our findings are summarized in Figures \ref{p3}-\ref{p2}.  They represent 
the flows in the three-dimensional S-T-U moduli space for different charge configurations.
In these figures, the black dot represents the $AdS_2$ fixed point, while the blue and red dots represent the $AdS_4$ 
fixed points \eqref{ads-I} and \eqref{ads-II}, which we denote by type 1 and type 2 $AdS_4$ fixed points, respectively.
We also took $a= 0.01$ in these plots.

First consider the case when $p^1 =p^2 =0$. Then, we find a flow connecting the associated 
$AdS_2$ fixed point to the $AdS_4$ fixed point \eqref{ads-II}, as depicted in Figure \ref{p3}.
When $p^3 =0$, we find a flow connecting the associated $AdS_2$ fixed point 
to the $AdS_4$ fixed point \eqref{ads-I}, as depicted in Figure \ref{p1p2}.
And finally, when  $p^2=0$, we find a flow connecting the associated $AdS_2$ fixed point 
to the $AdS_4$ fixed point \eqref{ads-II}, as depicted in Figure \ref{p2}.

\begin{figure}
    \centering
    \includegraphics[scale=0.35]{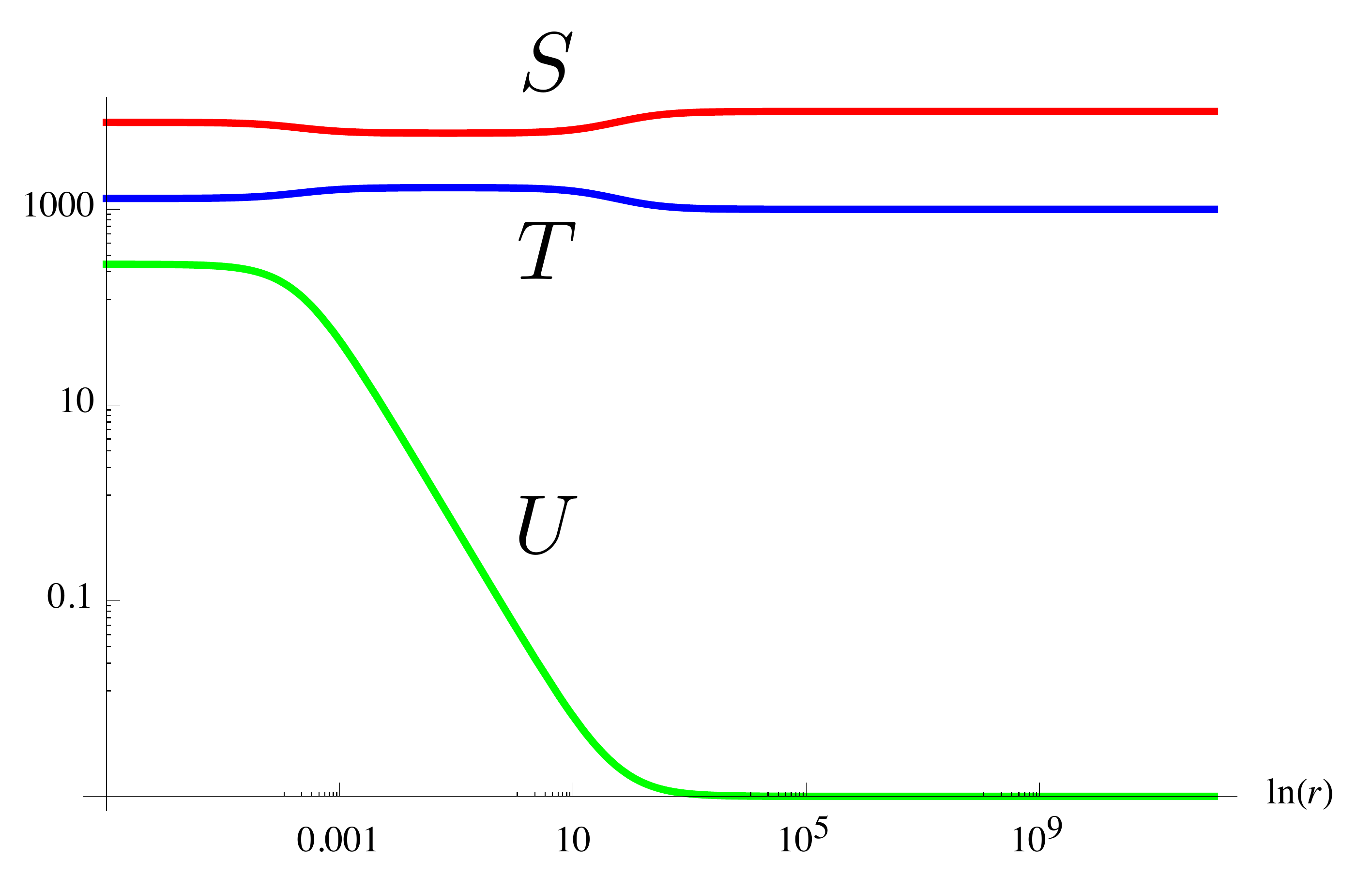}
  \caption{Flow in the perturbative chamber  $S>>T>U$ 
  $(q_0 =32, p^1 = 12, p^2 =2, p^3=0; h^0=h_1=1, h_2=10, h_3 =10^7)$, using $a=0.001$.\label{scalars_STU}}
\end{figure}

\begin{figure}
    \centering
    \includegraphics[scale=0.35]{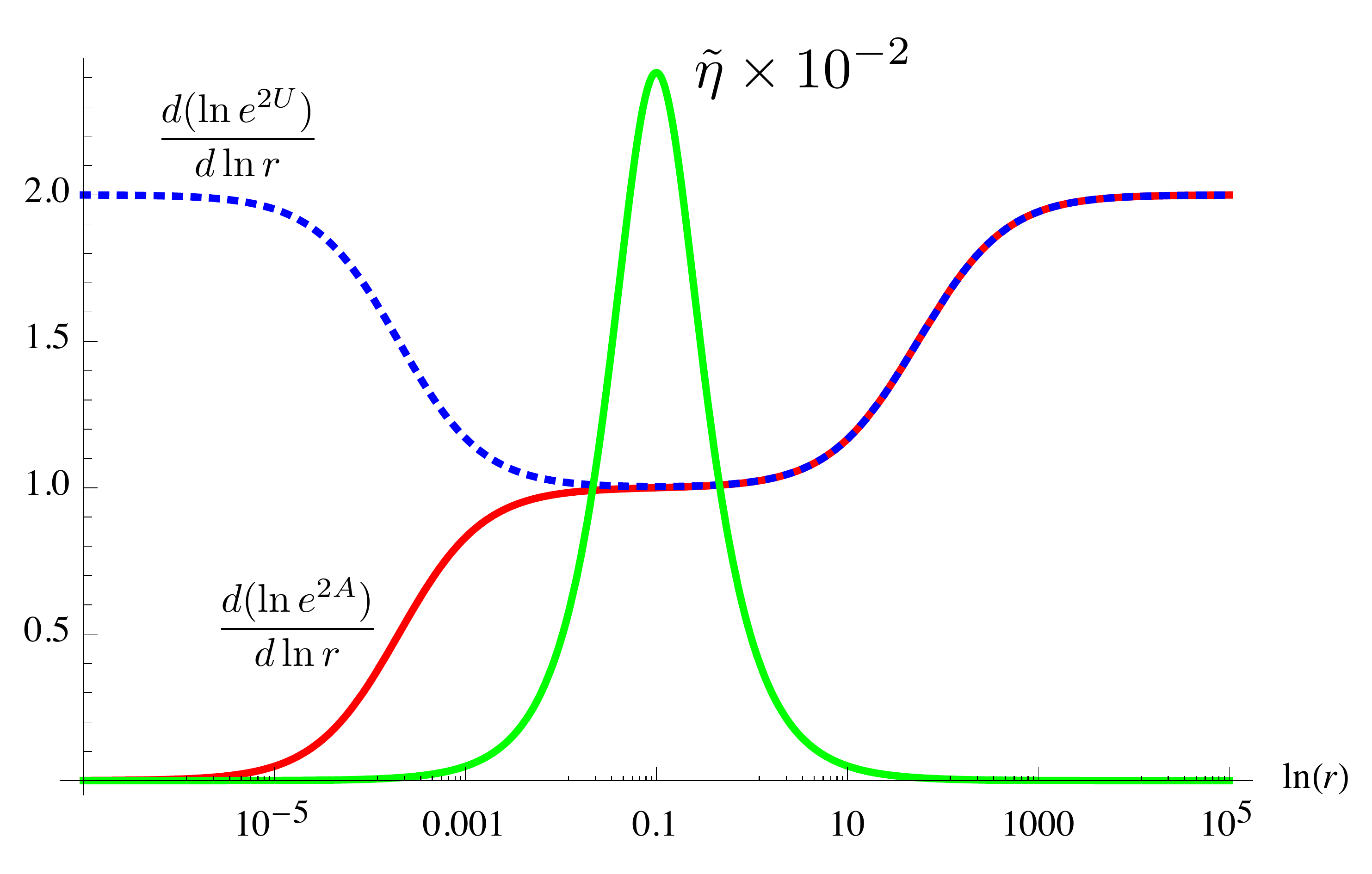}
  \caption{Metric behavior
  $(q_0 =32, p^1 = 12, p^2 =2, p^3=0; h^0=h_1=1, h_2=10, h_3 =10^7)$, using $a=0.001$. See \eqref{etatilde} for a definition of $\tilde{\eta}$.  \label{metric_STU}}
\end{figure}

When $p^3=0$, we can show that the flow of the scalar fields remains in the perturbative
chamber $S>>T>U$ when suitably choosing the charges and the fluxes. 
This is depicted in Figure \ref{scalars_STU}.
The behavior of the metric factors $e^{2U}$ and $e^{2A}$ is depicted in Figure \ref{metric_STU}. 
Note that, in a regime where 
$e^{2U}$ or $e^{2A}$ simply scale as powers, the quantities $\frac{d(\ln e^{2 U})}{d \ln r}$ and 
$\frac{d(\ln e^{2 A})}{d \ln r}$ give these powers, i.e.\ $\frac{d(\ln r^{\rho})}{d \ln r} = \rho$. From Figure \ref{metric_STU}
one reads off an intermediate scaling regime where both $e^{2U}$ and $e^{2A}$ roughly scale linearly in $r$. One might 
wonder whether this can be interpreted as a scaling regime with non-trivial dynamical critical exponent $z$ and hyperscaling violation parameter $\theta$. To answer his question we need to know the relation between the scalings of $e^{2U}$ and $e^{2A}$ on the one hand and $\theta$ and $z$ on the other hand. For 
\begin{equation}
e^{2U}\sim r^{2 \alpha} \quad , \quad e^{2 A}\sim r^{2\beta}
\end{equation}
one finds \cite{Bhattacharya:2012zu} 
\begin{eqnarray} \label{thetaz}
\theta = \frac{2 (\alpha-1)}{\alpha + \beta - 1} \quad , \quad z = \frac{2 \alpha - 1}{\alpha + \beta - 1}\ .
\end{eqnarray}
To answer the question whether there is an intermediate scaling regime we focus on 
\begin{eqnarray} \label{eta}
\eta = -\frac{\theta}{z} = -\frac{2 (\alpha-1)}{2 \alpha - 1}\ .
\end{eqnarray} 
If there was a scaling regime with particular values of $\theta$ and $z$, one would have to see a plateau when plotting 
\begin{eqnarray} \label{etatilde}
\tilde{\eta} = -\frac{2 \left(\frac{d(\ln e^{U})}{d \ln r}-1\right)}{2 \frac{d( \ln e^{U})}{d \ln r} - 1}\ .
\end{eqnarray} 
This, however, is not the case, as depicted in Figure \ref{metric_STU}. 

Let us next come to an analysis of the near-horizon geometry for the $(q_0, p^1, p^2)$ confi\-gu\-ra\-tion when 
$a$ is completely switched off.  It is straightforward to check that the following ansatz solves the flow equations
\eqref{first order system} in the limit $r \rightarrow 0$, 
\begin{eqnarray}
Y^0 \sim \sqrt{r} \;\;\;,\;\;\; Y^1, Y^2 \sim r^0 \;\;\;,\;\;\; Y^3 \sim r \;\;\;,\;\; e^{\psi} =r \;,
\end{eqnarray}
with $\gamma =0$.  This 
results in $e^{2A} \sim r^{1/2}$ and $e^{2U} \sim r^{3/2}$.  The associated line element
\begin{equation}
ds^2 = r^{1/2} \left[- r dt^2 + r^{-2} dr^2 +  dx^2 + dy^2 \right] \;,
\end{equation}
describes a so-called $\eta$-geometry \cite{Donos:2012yi}, namely 
\begin{equation}
ds^2 = {\tilde r}^{- \eta} \left[- {\tilde r}^{-2} dt^2 + l^2 \,  {\tilde r}^{-2} d {\tilde r}^2 +  dx^2 + dy^2 \right] \;,
\end{equation}
where $r = {\tilde r}^2 $, $l^2 = 4$ and $\eta = 1$. When turning on the regulator $a$, this $\eta$-geometry
gets modified into the $AdS_2 \times \mathbb{R}^2$-geometry discussed above for the $(q_0, p^1, p^2)$-system.

\begin{figure}
    \centering
    \includegraphics[scale=0.35]{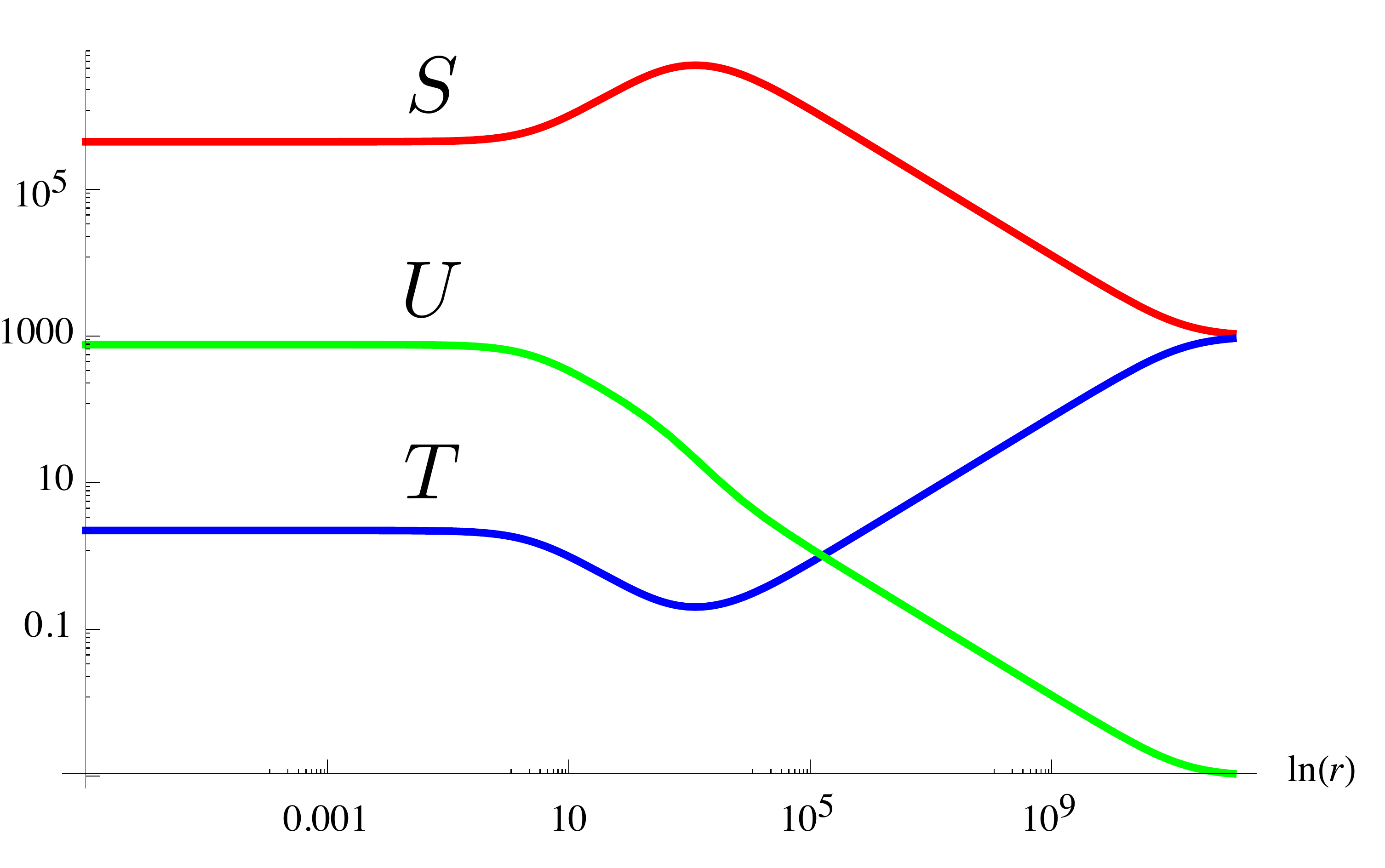}
  \caption{Flow in the purely magnetic case 
  $(q_0 =0, p^1 = 1.999.990, p^2 =10, p^3=-2; h^0=h_1=h_2=1, h_3 =10^6)$, using $a=0.001$.\label{scalars_magnetic}}
\end{figure}

\begin{figure}
    \centering
    \includegraphics[scale=0.35]{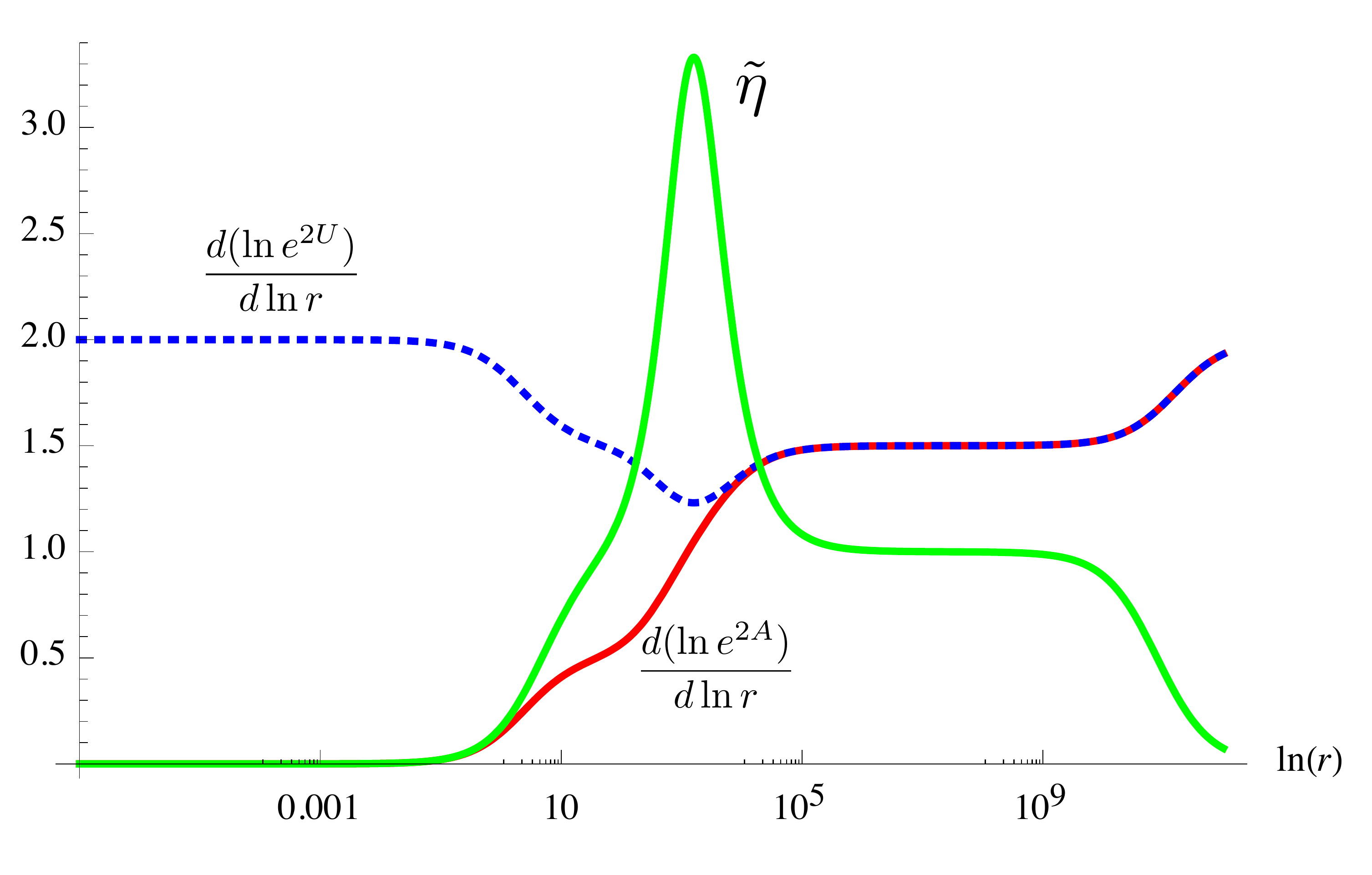}
  \caption{Metric behavior 
  $(q_0 =0, p^1 = 1.999.990, p^2 =10, p^3=-2; h^0=h_1=h_2=1, h_3 =10^6)$, using $a=0.001$. See \eqref{etatilde} for a definition of $\tilde{\eta}$.  \label{metric_magnetic}}
\end{figure}

Finally we would like to give an example of a flow which is purely magnetic, as this is the case that much of the earlier
literature focussed on, cf.\ \cite{Harrison:2012vy,Bhattacharya:2012zu}. In this case it is possible to find a scaling regime with $\eta \approx 1$. However, it does not stay all the way in the perturbative chamber, cf.\ Figures \ref{scalars_magnetic} and \ref{metric_magnetic}. One way to enforce staying in the perturbative chamber in the purely magnetic case would be to choose $p_1$ and $p_2$ of equal order (i.e.\ both of order $10^6$ in the example of Figures \ref{scalars_magnetic} and \ref{metric_magnetic}). However, in that case the $\eta$-scaling regime disappears and the plots look very similar to Figures \ref{scalars_STU} and \ref{metric_STU}. 

We would like to end with a comment on the $a \rightarrow 0$ limit. One might expect that in this limit the scaling regime of Figure \ref{metric_magnetic} extends more and more into the $AdS_2 \times \mathbb{R}^2$-region. To a small extend this indeed happens and the scaling regime also gets more extended to larger values of $r$ when decreasing the value of $a$. However, the effect of decreasing $a$ is actually surprisingly small. Changing $a$ from $10^{-1}$ to $10^{-7}$ cuts the $AdS_2 \times \mathbb{R}^2$-region only by a factor of about $10$. The reason for this small effect seems to be that the attractor values of the scalars $S,T$ and $U$ are getting larger for smaller values of $a$, counterbalancing the decrease of $a$ in the quantum correction $a U^3$ to the prepotential and, thus, preventing it from becoming negligible.

\vskip 5mm

\subsection*{Acknowledgements}
We would like to thank Gon\c{c}alo Marques Oliveira for very helpful discussions and 
Falk Ha\ss ler for his support concerning Mathematica. 
The work of G.L.C. is supported by the Center for Mathematical Analysis, Geometry
and Dynamical Systems (IST/Portugal), a unit of the LARSyS laboratory, as well as 
by Funda\c{c}\~{a}o para a Ci\^{e}ncia e a Tecnologia
(FCT/Portugal) through grants PTDC/MAT/119689/2010 and EXCL/MAT-GEO/0222/2012. 
The work of A.V.-O. is supported by Funda\c{c}\~{a}o para a Ci\^{e}ncia e a Tecnologia
(FCT/Portugal) through grant
SFRH/BD/64446/2009. The work of S.B.-D. and M.H.\ is supported by the Excellence Cluster 
``The Origin and the Structure of the Universe'' in Munich. The work of M.H. 
is also supported by the German Research Foundation (DFG) 
within the Emmy-Noether-Program (grant number: HA 3448/3-1).
This work is also supported by the COST action
MP1210 {\it The String Theory Universe}.


\appendix

\section{Homogeneity relations \label{homog-rel}}

The homogeneity relation $F(\lambda \, Y, \lambda^2 \, \Upsilon^2) = \lambda^2 \, F(Y, \Upsilon)$ implies
\begin{eqnarray}
2 F &=& Y^I F_I + 2 \Upsilon F_{\Upsilon} \;, \nonumber\\
Y^I F_{IJK} &=& - 2 \Upsilon F_{\Upsilon JK} \;, \nonumber\\
Y^I F_{I \Upsilon} &=& - 2 \Upsilon F_{\Upsilon \Upsilon} \;, \nonumber\\
F_{\Upsilon I} + Y^J F_{\Upsilon J I } &=& -2 \Upsilon F_{\Upsilon \Upsilon I} \;.
\label{hom-rel}
\end{eqnarray}

\section{Field combinations \label{superconf-free}}

The computation of the free energy \eqref{eq:F} makes use of various field combinations
that in \cite{Cardoso:2006xz} were computed for an $AdS_2 \times S^2$ geometry.  Here we adapt these
results to a background of the form \eqref{line-k}.  
Indices $i,j$ refer to the $AdS_2$ coordinates
$r,t$, whereas indices $\alpha,\beta$ refer to coordinates of $d\Omega^2_k$.  We obtain
for the field combinations considered in \cite{Cardoso:2006xz}, 
\begin{eqnarray}
  \label{eq:background-quantities}
  R &=& 2 \left( v_1^{-1} - k \, v_2^{-1} \right) \;,\nonumber\\
  f_i{}^j &=& [\ft12 v_1^{-1} - \ft14 (D + \ft13 R) -\ft{1}{32}  
    \vert w \vert^2]\,\delta_i{}^j \;, \nonumber\\
  f_\alpha{}^\beta &=& [ - \tfrac{k}{2} v_2^{-1} - \ft14 (D + \ft13 R)
  +\ft{1}{32} \vert w \vert^2 ]\,\delta_\alpha{}^\beta \;, \nonumber\\
  \mathcal{R}(M)_{ij}{}^{kl} &=& (D+\ft13 R)\,\delta_{ij}{}^{kl} \;,
  \nonumber \\  
  \mathcal{R}(M)_{\alpha\beta}{}^{\gamma\delta} &=& (D+\ft13
  R)\,\delta_{\alpha\beta}{}^{\gamma\delta} \;,   \nonumber \\    
   \mathcal{R}(M)_{i\alpha}{}^{j \beta} &=& \ft12 (D- \ft16
  R)\,\delta_i^j\,\delta_\alpha^\beta \;,   \nonumber \\    
  {\hat A} &=&  - 4 w^2 \;, \nonumber\\
  {\hat F}^-_{\underline{r}\underline{t}} &=&
  - 16 w (D
  + \ft13 R) \;,   \nonumber\\ 
  {\hat C} &=& 192 D^2 +   \ft{32}3  R^2 
  - 16 \vert w \vert^2 ( v_1^{-1}+ k \,v_2^{-1}) +2\vert w \vert^4 \;,
\end{eqnarray} 
where we recall that $k$ denotes the curvature of the two-dimensional space
with line element $d \Omega^2_k$.

The resulting field equations for $\tilde D$ and $\tilde \chi$ become
\begin{eqnarray}
  \label{eq:tilde-D-chi}
  \tilde D &=& 0\,, \nonumber \\
  \tilde \chi &=& -\frac{16 i}{\sqrt{-\Upsilon}} (\bar Y^IF_I
  -Y^I\bar F_I) - 256 i (F_\Upsilon-\bar F_\Upsilon)
  (k-\Xi^{-1}) \nonumber \\
  &&
  + 32 i \sqrt{-\Upsilon}
  \Big[F_{I\Upsilon} N^{IJ} 
   (\mathcal{Q}_J - \bar F_{JK}\mathcal{P}^K) - \mathrm{h.c.}\Big] \,.
\end{eqnarray}

\section{Numerical interpolation \label{numer}}

Here we outline a procedure to perform the numerical interpolation of the flow equations \eqref{first order system}.
It is convenient to recast these equations in terms of a radial coordinate $\tau$ defined by
\begin{equation}
 e^{\psi}\frac{\partial}{\partial r}= -\frac{\partial}{\partial \tau}.
\end{equation}
For the following it is important that the flow equations are autonomous (i.e.\ the right hand sides do not depend explicitly on the independent variable $\tau$) and they read
\begin{equation}\label{Moduli Flow}
 \dot Y^I= -N^{IK}\left(\bar{\hat Q}_K+ige^{2A}\bar{\hat h}_K\right),
\end{equation}
\begin{equation}\label{psidot}
  \dot\psi= -2g\Im\left[W(Y)\right] .
\end{equation}  
Here we set $\gamma =0$.
In terms of $\tau$, the attractor nature of the horizon becomes 
manifest since, as $\tau\rightarrow \infty$,  the moduli  flow towards an equilibrium state, i.e. they tend towards constant values $\tilde Y^I$. Since $e^{\psi}$ tends to zero when approaching an 
$AdS_2\times\mathbb{R}^2$ geometry, it follows from \eqref{psidot} that 
\begin{equation} \label{ImWneg}
2g\Im\left[W(\tilde Y)\right]<0.
\end{equation}
In order to find interpolating solutions we must figure out how to move away from this equilibrium configuration. To understand 
the possible deviations we need to linearize the system (\ref{Moduli Flow}) around the attractor point and study the eigenvalues 
of the Jacobian of the system. This is a rather difficult task, but it is possible to show that there always exists at least one
stable direction, i.e. a class of deformations that eventually evolves back towards equilibrium. This can be shown
as follows.

Consider a deviation of the form
\begin{equation}\label{perturbation}
Y^I(\tau)= \tilde Y^I+\epsilon v^I e^{\lambda\tau} \;,
\end{equation}
where $\epsilon\ll 1$ . To linear order in $\epsilon$,  Eq.  (\ref{Moduli Flow}) becomes
\begin{equation}\label{linear}
\lambda v^I=  (\tilde{\cal J}^{(1)})_{\;J}^{I}v^J+ (\tilde{\cal J}^{(2)})_{\; J}^{I}\bar v^J \;,
\end{equation}
where
\begin{equation}
  (\tilde{\cal J}^{(1)})_{\;J}^{I}= \left. N^{IK}\left(\bar F_{JKL}\left(p^L+ige^{2A}h^L\right)+ig\left(N\bar Y\right)_J\bar{\hat h}_K\right)\right|_{Y^I(\tau) =  \tilde Y^I}
\end{equation}
and
\begin{equation}
  (\tilde{\cal J}^{(2)})_{\; J}^{I}=\left.  ig\left(N Y\right)_{J}  N^{IK} \bar{\hat h}_K\right|_{Y^I(\tau) =  \tilde Y^I}
\end{equation}
are the two contributions from the Jacobian of \eqref{Moduli Flow}.
Then we can show that 
\begin{equation}
   \tilde K_{\bar I} (\tilde{\cal J}^{(1)})_{\;J}^{I}= -igW(\tilde Y)\tilde K_{\bar J} \;\;\;,\;\;\; \tilde K_{\bar I}  (\tilde{\cal J}^{(2)})_{\; J}^{I}= -igW(\tilde Y) \tilde K_J \;,
\end{equation}
where we used $K_{\bar I}=\partial_{\bar Y^I}K= -(NY)_I$ and $K_J=\partial_{Y^J}K= -(N\bar Y)_J$ and the tilde over $K_{\bar I}$ indicates that it is evaluated at the equilibrium values $\tilde Y$.
The last equation, when combined with (\ref{linear}), implies that
\begin{equation}
  \lambda= 2g\Im\left[W(\tilde Y)\right]\; ,
\end{equation}
which is negative due to \eqref{ImWneg}. Hence $Y(\tau)$ given in (\ref{perturbation}) approaches equilibrium as $\tau\rightarrow\infty$. 
The next step is to find the components of $v^I$. As a matter of fact we can only find the direction in moduli space in which 
$v^I$ points, but this is all we need, since $\epsilon$ can be used to tune the size of the vector.
Having determined the aforementioned direction, we can perform a numerical integration in order to find the flow line passing through
\begin{equation}
  Y(0)=\tilde Y^I+\epsilon v^I.
\end{equation}
on its way towards the horizon. Effectively, in the examples we looked at, the resulting flow coincides with the one that follows from the procedure outlined in sec.\ 3.1.2 of \cite{Barisch:2011ui}.\footnote{We would be happy to make our mathematica code available upon request.}

\bibliographystyle{JHEP}
\bibliography{references}

\providecommand{\href}[2]{#2}\begingroup\raggedright\begin{thebibliography}{10}

\bibitem{Goldstein:2009cv}
K.~Goldstein, S.~Kachru, S.~Prakash, and S.~P. Trivedi, {\it {Holography of
  Charged Dilaton Black Holes}},  {\em JHEP} {\bf 1008} (2010) 078,
  [\href{http://xxx.lanl.gov/abs/0911.3586}{{\tt 0911.3586}}].

\bibitem{Kachru:2008yh}
S.~Kachru, X.~Liu, and M.~Mulligan, {\it {Gravity Duals of Lifshitz-like Fixed
  Points}},  {\em Phys.Rev.} {\bf D78} (2008) 106005,
  [\href{http://xxx.lanl.gov/abs/0808.1725}{{\tt 0808.1725}}].

\bibitem{Hartnoll:2009sz}
S.~A. Hartnoll, {\it {Lectures on holographic methods for condensed matter
  physics}},  {\em Class.Quant.Grav.} {\bf 26} (2009) 224002,
  [\href{http://xxx.lanl.gov/abs/0903.3246}{{\tt 0903.3246}}].

\bibitem{Charmousis:2010zz}
C.~Charmousis, B.~Gouteraux, B.~S. Kim, E.~Kiritsis, and R.~Meyer, {\it
  {Effective Holographic Theories for low-temperature condensed matter
  systems}},  {\em JHEP} {\bf 1011} (2010) 151,
  [\href{http://xxx.lanl.gov/abs/1005.4690}{{\tt 1005.4690}}].

\bibitem{Copsey:2010ya}
K.~Copsey and R.~Mann, {\it {Pathologies in Asymptotically Lifshitz
  Spacetimes}},  {\em JHEP} {\bf 1103} (2011) 039,
  [\href{http://xxx.lanl.gov/abs/1011.3502}{{\tt 1011.3502}}].

\bibitem{Horowitz:2011gh}
G.~T. Horowitz and B.~Way, {\it {Lifshitz Singularities}},  {\em Phys.Rev.}
  {\bf D85} (2012) 046008, [\href{http://xxx.lanl.gov/abs/1111.1243}{{\tt
  1111.1243}}].

\bibitem{Blau:2009gd}
M.~Blau, J.~Hartong, and B.~Rollier, {\it {Geometry of Schrodinger Space-Times,
  Global Coordinates, and Harmonic Trapping}},  {\em JHEP} {\bf 0907} (2009)
  027, [\href{http://xxx.lanl.gov/abs/0904.3304}{{\tt 0904.3304}}].

\bibitem{Shaghoulian:2011aa}
E.~Shaghoulian, {\it {Holographic Entanglement Entropy and Fermi Surfaces}},
  {\em JHEP} {\bf 1205} (2012) 065,
  [\href{http://xxx.lanl.gov/abs/1112.2702}{{\tt 1112.2702}}].

\bibitem{Lei:2013apa}
Y.~Lei and S.~F. Ross, {\it {Extending the nonsingular hyperscaling violating
  spacetimes}},  \href{http://xxx.lanl.gov/abs/1310.5878}{{\tt 1310.5878}}.

\bibitem{Huijse:2011ef}
L.~Huijse, S.~Sachdev, and B.~Swingle, {\it {Hidden Fermi surfaces in
  compressible states of gauge-gravity duality}},  {\em Phys.Rev.} {\bf B85}
  (2012) 035121, [\href{http://xxx.lanl.gov/abs/1112.0573}{{\tt 1112.0573}}].

\bibitem{Hartnoll:2012wm}
S.~A. Hartnoll and E.~Shaghoulian, {\it {Spectral weight in holographic scaling
  geometries}},  {\em JHEP} {\bf 1207} (2012) 078,
  [\href{http://xxx.lanl.gov/abs/1203.4236}{{\tt 1203.4236}}].

\bibitem{Kundu:2012jn}
N.~Kundu, P.~Narayan, N.~Sircar, and S.~P. Trivedi, {\it {Entangled Dilaton
  Dyons}},  {\em JHEP} {\bf 1303} (2013) 155,
  [\href{http://xxx.lanl.gov/abs/1208.2008}{{\tt 1208.2008}}].

\bibitem{Donos:2012yi}
A.~Donos, J.~P. Gauntlett, and C.~Pantelidou, {\it {Semi-local quantum
  criticality in string/M-theory}},  {\em JHEP} {\bf 1303} (2013) 103,
  [\href{http://xxx.lanl.gov/abs/1212.1462}{{\tt 1212.1462}}].

\bibitem{Harrison:2012vy}
S.~Harrison, S.~Kachru, and H.~Wang, {\it {Resolving Lifshitz Horizons}},
  \href{http://xxx.lanl.gov/abs/1202.6635}{{\tt 1202.6635}}.

\bibitem{Bhattacharya:2012zu}
J.~Bhattacharya, S.~Cremonini, and A.~Sinkovics, {\it {On the IR completion of
  geometries with hyperscaling violation}},  {\em JHEP} {\bf 1302} (2013) 147,
  [\href{http://xxx.lanl.gov/abs/1208.1752}{{\tt 1208.1752}}].

\bibitem{Knodel:2013fua}
G.~Knodel and J.~T. Liu, {\it {Higher derivative corrections to Lifshitz
  backgrounds}},  {\em JHEP} {\bf 1310} (2013) 002,
  [\href{http://xxx.lanl.gov/abs/1305.3279}{{\tt 1305.3279}}].

\bibitem{Donos:2011qt}
A.~Donos, J.~P. Gauntlett, and C.~Pantelidou, {\it {Spatially modulated
  instabilities of magnetic black branes}},  {\em JHEP} {\bf 1201} (2012) 061,
  [\href{http://xxx.lanl.gov/abs/1109.0471}{{\tt 1109.0471}}].

\bibitem{Donos:2011pn}
A.~Donos, J.~P. Gauntlett, and C.~Pantelidou, {\it {Magnetic and Electric AdS
  Solutions in String- and M-Theory}},  {\em Class.Quant.Grav.} {\bf 29} (2012)
  194006, [\href{http://xxx.lanl.gov/abs/1112.4195}{{\tt 1112.4195}}].

\bibitem{Cremonini:2012ir}
S.~Cremonini and A.~Sinkovics, {\it {Spatially Modulated Instabilities of
  Geometries with Hyperscaling Violation}},
  \href{http://xxx.lanl.gov/abs/1212.4172}{{\tt 1212.4172}}.

\bibitem{Iizuka:2013ag}
N.~Iizuka and K.~Maeda, {\it {Stripe Instabilities of Geometries with
  Hyperscaling Violation}},  {\em Phys.Rev.} {\bf D87} (2013) 126006,
  [\href{http://xxx.lanl.gov/abs/1301.5677}{{\tt 1301.5677}}].

\bibitem{Donos:2013gda}
A.~Donos and J.~P. Gauntlett, {\it {Holographic charge density waves}},
  \href{http://xxx.lanl.gov/abs/1303.4398}{{\tt 1303.4398}}.

\bibitem{Cremonini:2013epa}
S.~Cremonini, {\it {Spatially Modulated Instabilities for Scaling Solutions at
  Finite Charge Density}},  \href{http://xxx.lanl.gov/abs/1310.3279}{{\tt
  1310.3279}}.

\bibitem{Cacciatori:2009iz}
S.~L. Cacciatori and D.~Klemm, {\it {Supersymmetric AdS(4) black holes and
  attractors}},  {\em JHEP} {\bf 1001} (2010) 085,
  [\href{http://xxx.lanl.gov/abs/0911.4926}{{\tt 0911.4926}}].

\bibitem{Dall'Agata:2010gj}
G.~Dall'Agata and A.~Gnecchi, {\it {Flow equations and attractors for black
  holes in N = 2 U(1) gauged supergravity}},  {\em JHEP} {\bf 1103} (2011) 037,
  [\href{http://xxx.lanl.gov/abs/1012.3756}{{\tt 1012.3756}}].

\bibitem{Hristov:2010ri}
K.~Hristov and S.~Vandoren, {\it {Static supersymmetric black holes in $AdS_4$
  with spherical symmetry}},  {\em JHEP} {\bf 1104} (2011) 047,
  [\href{http://xxx.lanl.gov/abs/1012.4314}{{\tt 1012.4314}}].

\bibitem{Barisch:2011ui}
S.~Barisch, G.~L.~Cardoso, M.~Haack, S.~Nampuri, and N.~A. Obers, {\it {Nernst
  branes in gauged supergravity}},  {\em JHEP} {\bf 1111} (2011) 090,
  [\href{http://xxx.lanl.gov/abs/1108.0296}{{\tt 1108.0296}}].

\bibitem{Sen:2005wa}
A.~Sen, {\it {Black hole entropy function and the attractor mechanism in higher
  derivative gravity}},  {\em JHEP} {\bf 0509} (2005) 038,
  [\href{http://xxx.lanl.gov/abs/hep-th/0506177}{{\tt hep-th/0506177}}].

\bibitem{Sahoo:2006rp}
B.~Sahoo and A.~Sen, {\it {Higher derivative corrections to non-supersymmetric
  extremal black holes in N=2 supergravity}},  {\em JHEP} {\bf 0609} (2006)
  029, [\href{http://xxx.lanl.gov/abs/hep-th/0603149}{{\tt hep-th/0603149}}].

\bibitem{Cardoso:2006xz}
G.~L.~Cardoso, B.~de~Wit, and S.~Mahapatra, {\it {Black hole entropy functions
  and attractor equations}},  {\em JHEP} {\bf 0703} (2007) 085,
  [\href{http://xxx.lanl.gov/abs/hep-th/0612225}{{\tt hep-th/0612225}}].

\bibitem{deWit:1979ug}
B.~de~Wit, J.~van Holten, and A.~Van~Proeyen, {\it {Transformation Rules of N=2
  Supergravity Multiplets}},  {\em Nucl.Phys.} {\bf B167} (1980) 186.

\bibitem{deWit:1980tn}
B.~de~Wit, J.~van Holten, and A.~Van~Proeyen, {\it {Structure of N=2
  Supergravity}},  {\em Nucl.Phys.} {\bf B184} (1981) 77.

\bibitem{deWit:1984pk}
B.~de~Wit and A.~Van~Proeyen, {\it {Potentials and Symmetries of General Gauged
  N=2 Supergravity: Yang-Mills Models}},  {\em Nucl.Phys.} {\bf B245} (1984)
  89.

\bibitem{deWit:1984px}
B.~de~Wit, P.~Lauwers, and A.~Van~Proeyen, {\it {Lagrangians of N=2
  Supergravity - Matter Systems}},  {\em Nucl.Phys.} {\bf B255} (1985) 569.

\bibitem{LopesCardoso:2000qm}
G.~L.~Cardoso, B.~de~Wit, J.~Kappeli, and T.~Mohaupt, {\it {Stationary BPS
  solutions in N=2 supergravity with $R^2$ interactions}},  {\em JHEP} {\bf
  0012} (2000) 019, [\href{http://xxx.lanl.gov/abs/hep-th/0009234}{{\tt
  hep-th/0009234}}].

\bibitem{Ceresole:2007wx}
A.~Ceresole and G.~Dall'Agata, {\it {Flow Equations for Non-BPS Extremal Black
  Holes}},  {\em JHEP} {\bf 0703} (2007) 110,
  [\href{http://xxx.lanl.gov/abs/hep-th/0702088}{{\tt hep-th/0702088}}].

\bibitem{Ferrara:1997tw}
S.~Ferrara, G.~W. Gibbons, and R.~Kallosh, {\it {Black holes and critical
  points in moduli space}},  {\em Nucl.Phys.} {\bf B500} (1997) 75--93,
  [\href{http://xxx.lanl.gov/abs/hep-th/9702103}{{\tt hep-th/9702103}}].

\bibitem{Ceresole:1995jg}
A.~Ceresole, R.~D'Auria, S.~Ferrara, and A.~Van~Proeyen, {\it {Duality
  transformations in supersymmetric Yang-Mills theories coupled to
  supergravity}},  {\em Nucl.Phys.} {\bf B444} (1995) 92--124,
  [\href{http://xxx.lanl.gov/abs/hep-th/9502072}{{\tt hep-th/9502072}}].

\bibitem{deWit:1996ag}
B.~de~Wit, {\it {N=2 symplectic reparametrizations in a chiral background}},
  {\em Fortsch.Phys.} {\bf 44} (1996) 529--538,
  [\href{http://xxx.lanl.gov/abs/hep-th/9603191}{{\tt hep-th/9603191}}].

\bibitem{Ooguri:2004zv}
H.~Ooguri, A.~Strominger, and C.~Vafa, {\it {Black hole attractors and the
  topological string}},  {\em Phys.Rev.} {\bf D70} (2004) 106007,
  [\href{http://xxx.lanl.gov/abs/hep-th/0405146}{{\tt hep-th/0405146}}].

\bibitem{Harvey:1995fq}
J.~A. Harvey and G.~W. Moore, {\it {Algebras, BPS states, and strings}},  {\em
  Nucl.Phys.} {\bf B463} (1996) 315--368,
  [\href{http://xxx.lanl.gov/abs/hep-th/9510182}{{\tt hep-th/9510182}}].

\bibitem{deWit:1996ix}
B.~de~Wit, {\it {N=2 electric - magnetic duality in a chiral background}},
  {\em Nucl.Phys.Proc.Suppl.} {\bf 49} (1996) 191--200,
  [\href{http://xxx.lanl.gov/abs/hep-th/9602060}{{\tt hep-th/9602060}}].

\bibitem{Dabholkar:2004yr}
A.~Dabholkar, {\it {Exact counting of black hole microstates}},  {\em
  Phys.Rev.Lett.} {\bf 94} (2005) 241301,
  [\href{http://xxx.lanl.gov/abs/hep-th/0409148}{{\tt hep-th/0409148}}].

\bibitem{Dabholkar:2004dq}
A.~Dabholkar, R.~Kallosh, and A.~Maloney, {\it {A Stringy cloak for a classical
  singularity}},  {\em JHEP} {\bf 0412} (2004) 059,
  [\href{http://xxx.lanl.gov/abs/hep-th/0410076}{{\tt hep-th/0410076}}].

\bibitem{Sen:2004dp}
A.~Sen, {\it {How does a fundamental string stretch its horizon?}},  {\em JHEP}
  {\bf 0505} (2005) 059, [\href{http://xxx.lanl.gov/abs/hep-th/0411255}{{\tt
  hep-th/0411255}}].

\bibitem{LopesCardoso:2004xf}
G.~L.~Cardoso, B.~de~Wit, J.~Kappeli, and T.~Mohaupt, {\it {Asymptotic
  degeneracy of dyonic N = 4 string states and black hole entropy}},  {\em
  JHEP} {\bf 0412} (2004) 075,
  [\href{http://xxx.lanl.gov/abs/hep-th/0412287}{{\tt hep-th/0412287}}].

\bibitem{LopesCardoso:1998wt}
G.~L.~Cardoso, B.~de~Wit, and T.~Mohaupt, {\it {Corrections to macroscopic
  supersymmetric black hole entropy}},  {\em Phys.Lett.} {\bf B451} (1999)
  309--316, [\href{http://xxx.lanl.gov/abs/hep-th/9812082}{{\tt
  hep-th/9812082}}].

\bibitem{BarischDick:2012gj}
S.~Barisch-Dick, G.~L.~Cardoso, M.~Haack, and S.~Nampuri, {\it {Extremal black
  brane solutions in five-dimensional gauged supergravity}},  {\em JHEP} {\bf
  1302} (2013) 103, [\href{http://xxx.lanl.gov/abs/1211.0832}{{\tt
  1211.0832}}].

\end{thebibliography}\endgroup

\end{document}